\documentclass[11pt, a4paper]{article}
\usepackage[utf8]{inputenc}
\usepackage{amsmath}
\usepackage{placeins}
\usepackage{amsthm}
\usepackage{mathtools}
\usepackage{braket}
\usepackage{amssymb}
\usepackage{geometry}
\usepackage{csquotes}
\usepackage{fancyhdr}
\usepackage{pgfplots}
\pgfplotsset{compat=1.18}
\usepackage{comment}
\usepackage[toc,page]{appendix}
\usepackage[english]{babel}
\usepackage{hyperref}
\usepackage[export]{adjustbox}
\usepackage{caption}
\usepackage[usestackEOL]{stackengine}
\usepackage{makecell}
\usepackage[boxsize=0.5em,aligntableaux=center]{ytableau}
\usepackage{amsfonts}
\usepackage{amscd}
\usepackage{enumerate}
\usepackage{booktabs}
\usepackage{physics}
\usepackage{graphicx}
\usepackage{subcaption}
\usepackage{url}
\usepackage{enumitem}

\allowdisplaybreaks 
\numberwithin{equation}{section}

\usepackage[textsize=tiny,color=green]{todonotes} 

\newtheoremstyle{named}{}{}{\itshape}{}{\bfseries}{.}{.5em}{\thmnote{#3 }#1}
\theoremstyle{named}

\def\title#1{\centerline{\LARGE\bf\Longstack{#1}}\vskip .5em}

\def\bea{\begin{eqnarray}}
\def\eea{\end{eqnarray}}

\newcommand{\green}{%
  \mathord{\raisebox{-1.3ex}{\includegraphics[height=4.5ex]{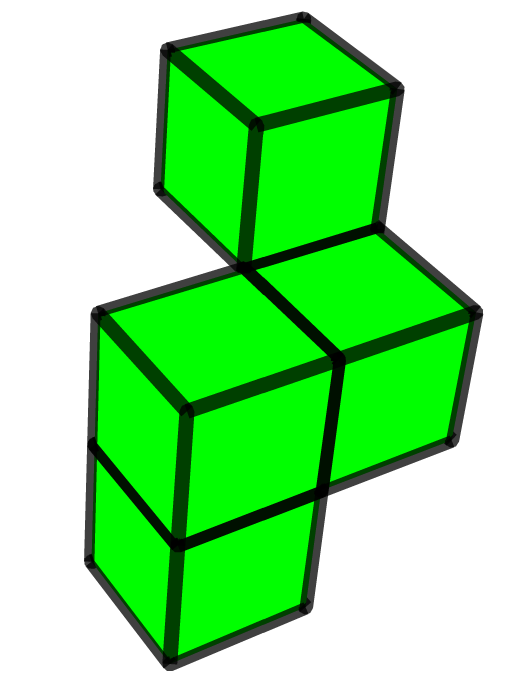}}}%
}
\newcommand{\purple}{%
  \mathord{\raisebox{-1.1ex}{\includegraphics[height=4.1ex]{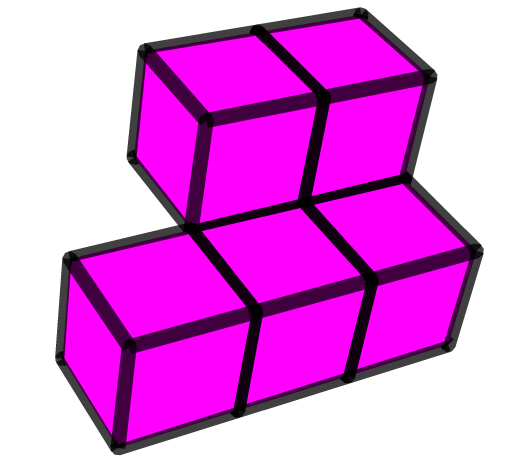}}}%
}
\newcommand{\yellow}{%
  \mathord{\raisebox{-1.3ex}{\includegraphics[height=4ex]{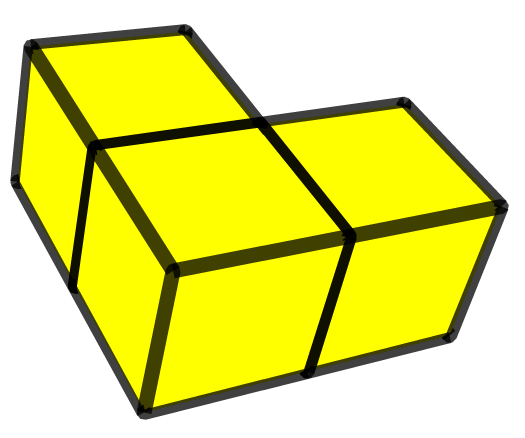}}}%
}
\newcommand{\cyan}{%
  \mathord{\raisebox{-1ex}{\includegraphics[height=4ex]{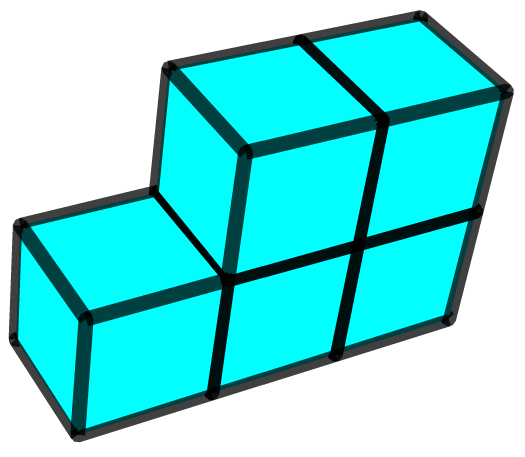}}}%
}
\newcommand{\orange}{%
  \mathord{\raisebox{-1ex}{\includegraphics[height=4ex]{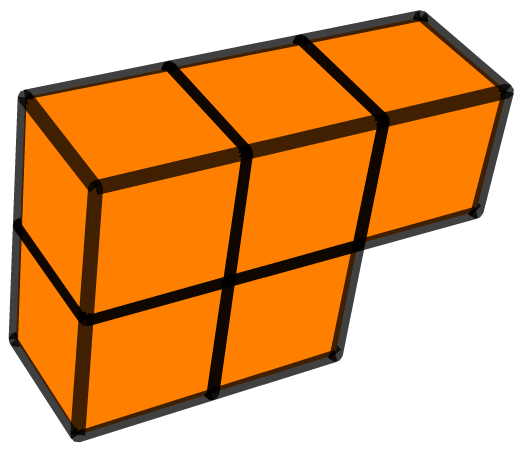}}}%
}

\newcommand{\origin}{\omega}

\newcommand\schr{Schr\" odinger }

\hypersetup{hidelinks}
\hypersetup{
  pdftitle={Quantum-mechanical bootstrap without inequalities: SYK bilinear spectrum},
  pdfauthor={Kok Hong Thong and David Vegh}
}
\begin{document}

\begin{flushright}
QMUL-PH-26-09\\
\end{flushright}

\bigskip
\bigskip

\title{
Quantum mechanical bootstrap without inequalities: \\  SYK bilinear spectrum
} 

\bigskip
\bigskip

\centerline{\bf Kok Hong Thong${}^{\dagger}$ and David Vegh${}^{\dagger \dagger}$}

\bigskip

\begin{center}

\small{
{ \it Centre for Theoretical Physics, Department of Physics and Astronomy \\
Queen Mary University of London, 327 Mile End Road, London E1 4NS, UK}}

\medskip
{\it Email:} ${}^{\dagger}$\texttt{k.h.thong@qmul.ac.uk}, ${}^{\dagger \dagger}$\texttt{d.vegh@qmul.ac.uk}

\bigskip
\bigskip
\centerline{ \it \today}

\end{center}

\begin{abstract}

We study a quantum mechanical system whose spectrum coincides with that of bilinear operators of the Sachdev--Ye--Kitaev model. The standard positivity-based quantum mechanical bootstrap is degenerate with respect to the boundary data: it does not distinguish the boundary conditions that select the SYK spectrum, and hence is insufficient to determine the eigenvalues. Instead, by considering fractional powers of operators, we obtain constraint equations that determine the spectrum without imposing positivity.
The resulting roots converge to exact eigenvalues as the truncation order increases. We call this the \textit{direct bootstrap}.
\end{abstract}

\tableofcontents


\section{Introduction}

The idea that consistency conditions alone can determine physical observables has a long history in theoretical physics.
In conformal field theory, the modern numerical bootstrap programme---initiated in \cite{Rattazzi:2008pe} and reviewed in \cite{Poland:2018epd}---has shown that crossing symmetry, unitarity, and the operator product expansion can be combined into rigorous, high-precision bounds on operator dimensions and OPE coefficients, famously giving the critical exponents of the 3d Ising model \cite{El-Showk:2012cjh,El-Showk:2014dwa}.
These results are obtained from the algebraic and positivity structure of the theory.

A natural question is whether the same philosophy can be applied to quantum mechanics.
In quantum mechanics, a basic consistency condition is the positivity of the inner product, $\expval{\mathcal{O}^\dagger \mathcal{O}} \geq 0$.
The quantum mechanical (QM) bootstrap implements these constraints systematically: recursion relations among correlators $\expval{X^m P^n}$ express higher moments in terms of lower ones and the energy, while positive semidefiniteness (PSD) of the bootstrap matrix $\mathcal{B}_{ij} = \expval{\mathcal{O}_i^\dagger \mathcal{O}_j}$ restricts the allowed values of the energy and the seed correlators.
Following the seminal work on matrix quantum mechanics and matrix models \cite{Lin:2020mme,Han:2020bkb}, related bootstrap constructions have since been developed for a broad range of quantum-mechanical systems without solving the Schr\"odinger equation.

The scope of the QM bootstrap has grown considerably in the past few years.
These include one-dimensional Schr\"odinger problems, such as anharmonic, double-well, and periodic potentials \cite{Berenstein:2021dyf,Berenstein:2021loy,Bhattacharya:2021btd,Aikawa:2021eai,Guo:2023gfi,Berenstein:2023ppj,Huang:2025sua},
systems with boundaries and nontrivial self-adjoint extensions \cite{Berenstein:2022ygg,Sword:2024gvv}, PT-symmetric and non-Hermitian Hamiltonians \cite{Khan:2022uyz,Khan:2024mhc}, higher-dimensional central potentials \cite{Lawrence:2025wyl}, and matrix-theory applications such as BFSS/D0-brane quantum mechanics \cite{Lin:2023owt}.
In most of these systems, the method works by imposing PSD constraints at increasing truncation order: the allowed region in the space of unknowns shrinks, and the energy eigenvalues are isolated as the boundaries of the allowed islands converge.
In some cases this convergence is extremely rapid, and the positivity constraints can even determine the spectrum exactly \cite{Berenstein:2021dyf,Berenstein:2023ppj,Aikawa:2025sib}.

When the system is defined on a finite interval and the wavefunctions obey more general boundary conditions, the analysis becomes more subtle.
The reason is that the momentum operator $P = -i\partial_x$ is not automatically self-adjoint on a bounded domain: its self-adjoint realization depends on the boundary conditions, and generic operators $\mathcal{O}$ need not map the domain of $H$ back into itself.
As was first observed in \cite{Berenstein:2022ygg}, this failure of domain preservation gives rise to \emph{domain anomalies}: boundary terms that appear in the commutator expectation values and modify the recursion relations.
These anomalies carry information about the boundary conditions, and their careful treatment is essential to any bootstrap on a bounded domain.

In this paper, we extend the QM bootstrap to a system defined on the interval $z\in[0,1]$.
The Hamiltonian is
\begin{equation}
    H = SZ(1-Z)S + \frac{(\tfrac12-\Delta)^2}{Z(1-Z)},
\end{equation}
where $S=i\partial_z$ and $Z=z$ are canonically conjugate, and $0<\Delta<\tfrac12$ is a real parameter.
This system arises as the Casimir operator of a folded string moving in rigid $\mathrm{AdS}_2$ with imaginary radius squared \cite{Vegh:2024uie,Vegh:2025kgx}.
For the appropriate self-adjoint extension, its spectrum coincides exactly with the spectrum of bilinear operators in the Sachdev--Ye--Kitaev (SYK) model \cite{Sachdev:1992fk,Kitaev:2015sim,Polchinski:2016xgd,Maldacena:2016hyu, Kitaev:2017awl}, with $\Delta = 1/q$ the conformal dimension of the Majorana fermions in the infrared limit of the $q$-body SYK model.
The bilinear operator spectrum---determined by the kernel eigenvalue equations $k_f(h) = 1$ (fermionic) and $k_b(h) = 1$ (bosonic)---is a central quantity in SYK physics: it controls the four-point function, the operator product expansion in the conformal limit, and the approach \cite{Maldacena:2016hyu,Polchinski:2016xgd,Gross:2017hcz} to quantum chaos \cite{Maldacena:2015waa}. The \schr equation is a general Legendre equation, and its solutions can be expressed using associated Legendre functions.

Our main finding is that, for this system, \emph{positivity bounds from the standard QM bootstrap are not sufficient to determine the eigenvalues}.
The reason is structural: anomalous constraints from integer-powered operators  cannot distinguish between different boundary conditions, so the positivity islands extend over a continuous range of energies.
This stands in contrast to previous applications of the QM bootstrap, where PSD constraints alone suffice to isolate the spectrum.
To overcome this limitation, we extend the operator basis from integer powers of the position operator to \emph{fractional} powers:
\begin{equation}
    \mathcal{O}_{\sigma,\zeta,\xi} = S^\sigma Z^\zeta (1-Z)^\xi, \qquad \sigma \in \mathbb{Z}_{\geq 0}, \quad \zeta,\xi \in \mathbb{R}.
\end{equation}
Because the recursion relations preserve the fractional parts of $\zeta$ and $\xi$, correlators split into distinct operator families labelled by a common fractional offset $\origin$.
The anomaly-free sector within each family closes on three unknowns (the energy and two correlators), but is insensitive to the boundary conditions.
The boundary information enters through operators with finite, nonzero domain anomalies, which produce exact constraint equations---one from each of three special operator families, $\origin \in \{0,\, 2\Delta,\, 4\Delta - 1\}$.
Crucially, each family probes the boundary conditions differently.
Once related by cross-family Taylor expansions, their anomalous constraints become non-degenerate and close into a system of three equations for three unknowns.
No positivity input is required: the spectrum is determined directly from the equality constraints.

We demonstrate the method explicitly for $\Delta = 1/4$ and $\Delta = 1/6$.
In both cases, our {\it direct bootstrap} method converges to the exact spectrum as the Taylor-expansion truncation order $N$ increases, with power-law convergence rates that we bound analytically.

The paper is organized as follows.
In Section \ref{sec:2}, we introduce the Hamiltonian, specify the self-adjoint extension corresponding to SYK, and review the exact spectrum.
In Section \ref{sec:3}, we develop the direct bootstrap: Section \ref{sec:3.1} discusses dressed anomalies and their separation into a
boundary/domain anomaly and a bulk correction; Section \ref{sec:3.2} constructs the anomaly-free correlator-cone and shows that it closes on three unknowns; and Section \ref{sec:3.3} extracts the constraint equations from the anomalous sector.
In Section \ref{sec:4}, we present the numerical results and convergence analysis.
We conclude with a discussion of the implications and possible future directions.


\section{The model}
\label{sec:2}

\subsection{The Hamiltonian}
\label{sec:2.1}

We consider a quantum-mechanical system with coordinate $z\in[0,1]$.
The Hilbert space is $\mathcal H=L^2([0,1],dz)$, with inner product
\begin{equation}
\label{eq:innerproduct}
    \braket{\phi}{\varphi}_0 \coloneq \int_0^1 dz\, \phi(z)^* \varphi(z),
    \qquad \forall\, \phi,\varphi\in\mathcal H.
\end{equation}
The Hamiltonian of interest is
\begin{equation}
\label{eq:Hamiltonian}
    H = SZ(1-Z)S + \frac{\left(\frac12-\Delta\right)^2}{Z(1-Z)},
\end{equation}
where $S$ and $Z$ are conjugate operators satisfying
\begin{equation}
    S=i\hbar \partial_z,\qquad Z=z,\qquad [S,Z]=i\hbar.
\end{equation}
We set $\hbar=1$ henceforth.

This Hamiltonian arises as the Casimir operator of a folded string moving in rigid $\mathrm{AdS}_2$ with imaginary radius squared \cite{Vegh:2024uie,Vegh:2025kgx}.
For the self-adjoint extension specified in Section \ref{sec:2.2}, its spectrum coincides with the spectrum of SYK bilinear operators, with $\Delta=1/q$ the conformal dimension of the Majorana fermions in the infrared limit of the $q$-body SYK model \cite{Maldacena:2016hyu}.

For bootstrap purposes, it is convenient to remove the negative powers of $Z$ in \eqref{eq:Hamiltonian}.
In the $z$-basis, the time-independent Schr\"odinger equation is
\begin{equation}
    E\varphi(z)
    =
    \frac{\left(\frac12-\Delta\right)^2}{z(1-z)}\varphi(z)
    +(2z-1)\varphi'(z)
    +z(z-1)\varphi''(z).
\end{equation}
Now define
\begin{equation}
\label{eq:transform}
    \varphi(z)=[z(1-z)]^a\psi(z),
    \qquad a\in\mathbb R,
\end{equation}
which gives
\begin{equation}
    E\psi(z)
    =
    \left[
    \frac{a^2-\left(\frac12-\Delta\right)^2}{z(z-1)}
    +2a(2a+1)
    \right]\psi(z)
    +(2a+1)(2z-1)\psi'(z)
    +z(z-1)\psi''(z).
\end{equation}
Choosing
\begin{equation}
    a=\pm\left(\frac12-\Delta\right)
\end{equation}
removes the term with negative powers of $z$.
The two choices are related by the $\Delta\to 1-\Delta$ symmetry of the model, so we take
\begin{equation}
    a=-\frac12+\Delta.
\end{equation}
Then the equation simplifies to
\begin{equation}
\label{eq:DE}
    \boxed{
    E_a\psi(z)
    =
    2\Delta(2z-1)\psi'(z)+z(z-1)\psi''(z)
    \eqcolon H_a\psi(z),
    }
\end{equation}
where
\begin{equation}
    E_a \coloneq E+2\Delta(1-2\Delta),
    \qquad
    H_a=-i\,2\Delta(2Z-1)S-Z(Z-1)S^2.
\end{equation}
The transformed Hilbert space is $\mathcal{H}_a=L^2([0,1],w \, dz)$, with inner product
\begin{equation}
\label{eq:innerproduct_a}
    \langle \psi_1|\psi_2\rangle_a
    \coloneq
    \int_0^1 dz\,w\,\psi_1(z)^*\psi_2(z),
    \qquad
    \forall \psi_1,\psi_2\in\mathcal{ H}_a .
\end{equation}
where $w = [z(1-z)]^{2\Delta-1}$. 
The map $\psi\mapsto\varphi=w^{1/2}\psi$ is unitary from $\mathcal H_a$ to the original Hilbert space $\mathcal H$.
Equation \eqref{eq:DE} is a special case of the hypergeometric differential equation.
It reduces to the interval model bootstrapped in \cite{Sword:2024gvv} when $\Delta=\tfrac12$.

Because the system is defined on a finite interval, the choice of domain is nontrivial.
In particular, $S=i\partial_z$ is not self-adjoint on the naive domain, and the relevant self-adjoint realization of $H_a$ is fixed by boundary conditions at $z=0,1$ \cite{Al-Hashimi:2020qvi,Sword:2024gvv}.
For now we regard $H_a$ as the formal differential operator in \eqref{eq:DE} acting on $\mathcal{H}_a$.
The self-adjoint extension relevant for the SYK spectrum will be specified below.

\subsection{The SYK bilinear spectrum}
\label{sec:2.2}

Equation \eqref{eq:DE} is exactly solvable, and for an appropriate choice of boundary conditions its spectrum coincides with that of the SYK bilinear operators \cite{Vegh:2024uie,Vegh:2025kgx}.
This makes the model an ideal testing ground for the quantum-mechanical bootstrap.

Near the endpoints, the transformed wavefunction behaves as
\begin{equation}
\label{eq:boundaryconditions}
    \psi(z)=
    \begin{cases}
        c_A+c_B z^{1-2\Delta}, & z\to 0,\\[4pt]
        \tilde c_A+\tilde c_B (1-z)^{1-2\Delta}, & z\to 1.
    \end{cases}
\end{equation}
These are fractional-power Robin boundary conditions.
A parity-preserving one-parameter family of self-adjoint extensions is obtained by imposing $c_B = r c_A$ and $\tilde{c}_B = r \tilde{c}_A$, with $r \in \mathbb{R}$ \cite{Vegh:2024uie,Vegh:2025kgx}.
The SYK bilinear spectrum corresponds to the particular value
\begin{equation}
\label{eq:boundary}
    r=
    \frac{1-\Delta}{\Delta}.
\end{equation}

The exact solutions of \eqref{eq:DE} can be written as
\begin{equation}
    \psi(z)
    =
    [z(1-z)]^{-a}
    \left[
        c_1\,P_{h-1}^{\,2\Delta-1}(2z-1)
        +
        c_2\,Q_{h-1}^{\,2\Delta-1}(2z-1)
    \right],
\end{equation}
where $c_1,c_2\in\mathbb C$, $P$ and $Q$ are associated Legendre functions, and
\begin{equation}
    h \coloneq \frac{1+\sqrt{1+4E}}{2}.
\end{equation}
Imposing \eqref{eq:boundaryconditions}--\eqref{eq:boundary}, the allowed values of $h$ are determined by
\begin{equation}
    k_f(h)=1
\end{equation}
for antisymmetric wavefunctions and by
\begin{equation}
    k_b(h)=1
\end{equation}
for symmetric wavefunctions, where
\begin{align}
    k_f(h)
    &=
    \frac{\Gamma(3-2\Delta)\bigl(\sin(2\pi\Delta)-\sin(\pi h)\bigr)\Gamma(2\Delta-h)\Gamma(h+2\Delta-1)}
    {\pi\,\Gamma(2\Delta+1)},\\
    k_b(h)
    &=
    -\frac{\pi\,\Gamma(2\Delta+1)\bigl(\sin(\pi h)-\sin(2\pi\Delta)\bigr)\csc\!\bigl(\pi(h-2\Delta)\bigr)\csc\!\bigl(\pi(2\Delta+h)\bigr)}
    {\Gamma(3-2\Delta)\Gamma(2\Delta-h)\Gamma(h+2\Delta-1)}.
\end{align}
These are precisely the equations that determine the fermionic and bosonic bilinear spectra in SYK.
The remainder of the paper shows how to recover this spectrum bootstrap-wise, without using the analytic form of the wavefunctions.

\section{Direct bootstrap}
\label{sec:3}

\subsection{Boundary conditions and anomalies}
\label{sec:3.1}
In order to simplify the calculations, the direct bootstrap will be formulated in terms of the dressed correlators 
\begin{equation}
\label{eq:dressed}
    \expval{\mathcal{O}} \coloneq \int_0^1 dz  \, \psi^* \mathcal{O} \psi. 
\end{equation}
This is not a correlator of $\mathcal{O}$ in $\mathcal{H}_a$ due to the absence of weight $w$ in the integrand.
Instead,
\begin{equation}
    \expval{\mathcal{O}} = \expval{w^{-1}\mathcal{O}}_a,
\end{equation}
so every dressed correlator is equivalently a correlator in $\mathcal{H}_a$ of the transformed insertion $w^{-1} \mathcal{O}$. 
Henceforth, we shall refer to the correlators as the dressed correlators, $\expval{\mathcal{O}}$.
The goal of \emph{direct bootstrap} is to extract constraint equations for $E_a$ from commutator recursion relations of the form
\begin{equation}
\label{eq:directbootstrap_master}
    \boxed{\langle[H_a,\mathcal{O}]\rangle  = \mathcal{A}(\mathcal{O}),}
\end{equation}
where $\mathcal{A}$ is the \emph{dressed anomaly}.
The dressed anomaly is related to the domain anomaly via
\begin{align}
    \mathcal{A}(\mathcal{O}) =&\expval{w^{-1}[H_a,\mathcal{O}]}_a \\
    =& \expval{[H_a,w^{-1}\mathcal{O}]}_a-\expval{[H_a,w^{-1}]\mathcal{O}}_a, \label{eq:trueanomaly}
\end{align}
where the first term on the RHS is the dressed domain anomaly of $\mathcal{O}$, or equivalently the domain anomaly of $w^{-1} \mathcal{O}$.
Domain anomalies arise when $w^{-1}\mathcal{O}$ does not leave the domain of $H_a$ invariant \cite{Berenstein:2022ygg}; that is, when $w^{-1}\mathcal{O} \ket{\psi} \notin D(H_a)$ for $\ket{\psi} \in D(w^{-1}\mathcal{O}) \cap D(H_a)$.
In that case, the formal commutator need not vanish even in an energy eigenstate.
The dressed anomaly can be calculated in a similar way as the domain anomaly, via
\begin{align}
\label{eq:anomaly1}
    \mathcal{A}(\mathcal{O}) &=\langle[H_a,\mathcal{O}]\rangle   \\
    &= \bra{H_a\psi}\ket{\mathcal{O}\psi} - \bra{\psi}\ket{\mathcal{O}H_a\psi}
    + \bra{\psi}\ket{H_a \mathcal{O} \psi}-\bra{H_a \psi} \ket{O \psi}\\ 
    &= \bra{\psi}\ket{H_a \mathcal{O} \psi}-\bra{H_a \psi} \ket{\mathcal{O} \psi}.
\end{align}
Here the bra-ket notation refers to the unweighted integral in \eqref{eq:dressed}, not the weighted Hilbert-space inner product.

We choose operators and the corresponding correlators to be
\begin{equation}
    \boxed{\mathcal{O}_{\sigma, \zeta, \xi}\coloneq S^\sigma Z^\zeta (1-Z)^\xi, \qquad f_{\sigma, \zeta, \xi} \coloneq \expval{O_{\sigma, \zeta, \xi}},}
\end{equation}
where $\sigma \in \mathbb{Z}_{\geq 0}$ and $\zeta, \xi \in \mathbb{R}$. 
This is similar to the operators used in previous QM bootstraps, except that here we have extended the powers of the position operators to real values.  
By definition, the correlators satisfy the ``expansion'' recursion relation on the $\zeta,\xi$-plane,
\begin{equation}
\label{eq:recursion4}
    (\yellow)_{\sigma, \zeta, \xi}: 0=-f_{\sigma, \zeta, \xi} + f_{\sigma, 1+\zeta, \xi} + f_{\sigma, \zeta, 1+\xi}.
\end{equation}
We represent the correlators and recursion relations diagrammatically as a three-dimensional lattice with axes $(\sigma,\zeta,\xi)$, as shown in Figure \ref{fig:cone}.
We take $\yellow$ to mean either the recursion relation or the RHS of \eqref{eq:recursion4}, depending on the context.
Then, the dressed anomaly can be computed to be (see Appendix \ref{appendix:A})
\begin{align}
\label{eq:anomaly2}
    \mathcal{A}(\mathcal{O}_{\sigma,\zeta,\xi}) &= i^\sigma \left[\gamma (\beta' \alpha-\beta\alpha') + (2\Delta-1) \gamma' \beta\alpha \right]_\epsilon^{1-\epsilon} -2(2\sigma+1)(2\Delta-1) f_{\sigma,\zeta,\xi} \nonumber \\ 
    &\quad +2i(2\Delta-1)f_{\sigma+1,\zeta,\xi}  -4i (2\Delta-1)f_{\sigma+1,\zeta+1,\xi}, 
\end{align}
where $\alpha(z) = \partial^{\sigma}_z \!\left[z^\zeta(1-z)^\xi\psi(z)\right]$, $\beta(z) = \psi(z)^*$ and $\gamma(z) = z(1-z)$.
The dressed anomaly has a boundary term 
\begin{equation}
\label{eq:anomaly3}
    \mathcal{A}_{\partial \Omega}(\mathcal{O}_{\sigma,\zeta,\xi})\coloneq i^\sigma \left[\gamma (\beta' \alpha-\beta\alpha') + (2\Delta-1) \gamma' \beta\alpha \right]_\epsilon^{1-\epsilon},
\end{equation}
and a bulk term 
\begin{equation}
    \mathcal{A}_\Omega(\mathcal{O}_{\sigma,\zeta,\xi})\coloneq - 2(2\sigma+1)(2\Delta-1) f_{\sigma,\zeta,\xi} +2i(2\Delta-1)f_{\sigma+1,\zeta,\xi}  -4i (2\Delta-1)f_{\sigma+1,\zeta+1,\xi},
\end{equation}
where $z \in \Omega = [\epsilon, 1-\epsilon]$.
The boundary term is the dressed domain anomaly, $\expval{[H_a,w^{-1}\mathcal{O}]}_a$ in \eqref{eq:trueanomaly}. 
It contains the dependence on the fractional Robin data, but must be handled carefully because of endpoint divergences.
Since 
$\expval{[H_a,\mathcal{O}_{\sigma,\zeta,\xi}]}= \mathcal{A}(\mathcal{O}_{\sigma,\zeta,\xi})$, the bulk term simply extends the recursion relations for the correlators.
As such, we henceforth refer to $\mathcal{A}_{\partial \Omega}$ as the anomaly, unless otherwise specified.
In Section \ref{sec:3.2}, we first generate the correlator cone for anomaly-free operators, and then extract constraints from the anomalous operators in Section \ref{sec:3.3}.

\subsection{Generating the anomaly-free correlator-cone}
\label{sec:3.2}

\begin{figure}
    \centering
    \includegraphics[width=0.45\linewidth]{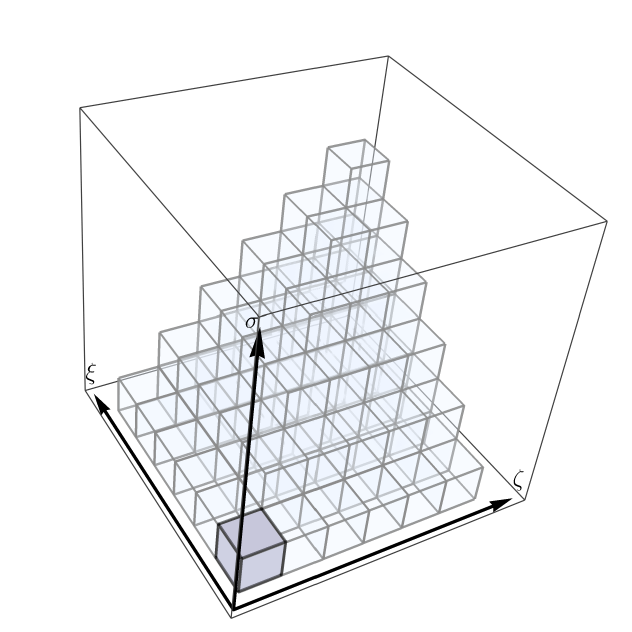}
    \caption{Anomaly-free correlator-cone for a fixed operator family labelled by $\origin$, defined by $\zeta,\xi>\sigma\geq 0$ in the $\epsilon\to 0$ limit. The gray block marks the reference point $(\sigma,\zeta,\xi)=(0,\origin,\origin)$.}
    \label{fig:cone}
\end{figure}
For the correspondence to SYK, we set $0< \Delta < 1/2$.
Then, taking $\epsilon \rightarrow 0$,  $\mathcal{A}_{\partial \Omega}(\mathcal{O}_{\sigma,\zeta,\xi}) = 0$ for $\zeta, \, \xi > \sigma \geq 0$, which defines the \emph{anomaly-free} correlator-cone shown in Figure \ref{fig:cone}.
Had we instead chosen a simpler operator basis such as $\mathcal{O}_{\sigma, \zeta}= S^\sigma Z^\zeta$, as in a previous QM bootstrap \cite{Sword:2024gvv}, only correlators with $\sigma = 0$ would have finite anomalies; for $\sigma > 0$, $\mathcal{A}_{\partial \Omega} \rightarrow \infty$ from the $z = 1$ boundary.
The inclusion of both $Z^\zeta$ and $(1-Z)^\xi$ is therefore essential here.

Because $H_a$ contains only integer powers of $Z$ and $(1-Z)$, the recursion relations preserve the fractional parts of $\zeta$ and $\xi$.
Hence, the correlators split into different operator families that are not related by \eqref{eq:directbootstrap_master}.
Accordingly, we write $\zeta \rightarrow \zeta + \origin$, $\xi \rightarrow \xi + \origin$ with $\zeta, \xi \in \mathbb{Z}_{\ge 0}$, and label operator families by the common fractional offset $\origin$.
More generally, the families are labelled by two fractional offsets; in this work we restrict to the diagonal subfamilies with a common offset $\origin$, which are sufficient for the symmetric and antisymmetric
sectors considered below.

To generate the anomaly-free cone, the steps are as follows:
\begin{enumerate}
    \item Generate the ``correlator-wall'' $f_{\sigma, \zeta+\origin, \origin+1}$, where $\sigma = 0, 1$ and $\zeta \geq 0$. 
    This can be achieved by finding two independent recursion relations along the wall.
    \item Use the expansion recursion, $\yellow$ \eqref{eq:recursion4}, to generate any $f_{\sigma, \zeta+\origin, \xi+\origin}$ for $\sigma = 0, 1$ (the first two floors).
    \item Find a recursion relation that generates $f_{\sigma, \zeta+\origin, \xi+\origin}$ for $\sigma \geq 2$ (the upper floors) in terms of the $f_{\sigma, \zeta+\origin, \xi+\origin}$ for $\sigma < 2$.
\end{enumerate}

\textbf{Step 1. Generating the correlator-wall $f_{\sigma, \zeta+\origin, \origin+1}$.} 
As in other QM bootstrap constructions, we begin with two independent recursion relations:
\begin{equation}
    \label{eq:recursion_master1}
    \expval{[H_a,\mathcal{O}_{\sigma,\zeta+\origin,\xi+\origin}]}- \mathcal{A}(\mathcal{O}_{\sigma,\zeta+\origin,\xi+\origin}) = 0, 
\end{equation}
\begin{equation}
    \label{eq:recursion_master2}
    \expval{H_a\mathcal{O}_{\sigma,\zeta+\origin,\xi+\origin}} - E_a\expval{\mathcal{O}_{\sigma,\zeta+\origin,\xi+\origin}}- \mathcal{A}(\mathcal{O}_{\sigma,\zeta+\origin,\xi+\origin}) = 0.
\end{equation}
Using standard commutation relations, \eqref{eq:recursion_master1} and \eqref{eq:recursion_master2} become
\begin{align}
\label{eq:recursion2}
    (\purple)_{\sigma, \zeta+\origin, \xi+\origin}: 0 =& \ g_{11} f_{\sigma ,\zeta +\origin,\xi +\origin-1} +g_{12} f_{\sigma ,\zeta +\origin-1,\xi +\origin-1}+g_{13} f_{\sigma +1,\zeta +\origin,\xi +\origin}\nonumber \\ 
    &+g_{14} f_{\sigma ,\zeta +\origin+1,\xi +\origin-1}+g_{15} f_{\sigma +1,\zeta +\origin+1,\xi +\origin}-\mathcal{A}_{\partial \Omega}(\mathcal{O}_{\sigma,\zeta+\origin,\xi+\origin}),
\end{align}
\begin{align}
\label{eq:recursion3}
    (\green)_{\sigma, \zeta+\origin, \xi+\origin}: 0=& \ g_{21} f_{\sigma ,\zeta +\origin,\xi +\origin}+g_{22}f_{\sigma +1,\zeta +\origin,\xi +\origin}+ g_{23} f_{\sigma +1,\zeta +\origin+1,\xi +\origin}  \nonumber \\ 
    &+g_{24}f_{\sigma +2,\zeta +\origin+1,\xi +\origin+1}-\mathcal{A}_{\partial \Omega}(\mathcal{O}_{\sigma,\zeta+\origin,\xi+\origin}) , 
\end{align}
where the coefficients are given explicitly in Appendix \ref{appendix:B} to prevent clutter.
If we had used $H$ instead of $H_a$, the negative powers of $Z(1-Z)$ would produce recursions that couple a large number of correlators, making the system difficult to close.

\begin{figure}
     \centering
     \begin{subfigure}[b]{0.49\textwidth}
         \centering
         \includegraphics[width=\textwidth]{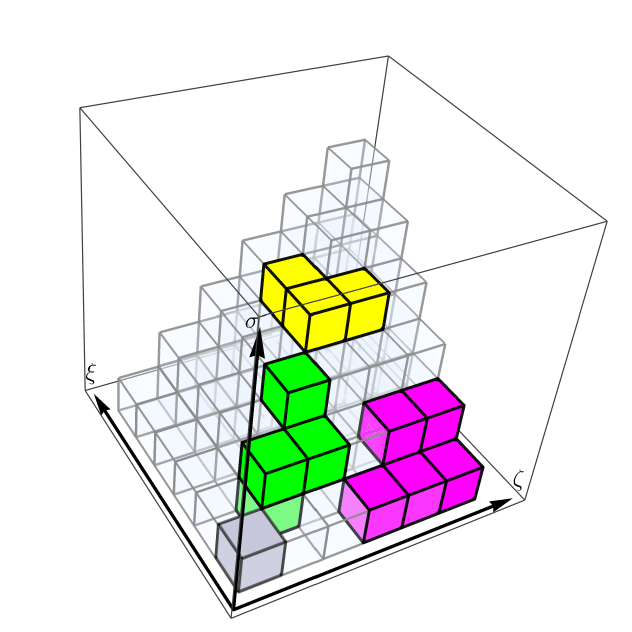}
         \caption{The basic recursion relations on the anomaly-free cone: the expansion recursion $\yellow$ \eqref{eq:recursion4}, the $\expval{H_a\mathcal{O}_{\sigma,\zeta+\origin,\xi+\origin}}-E_a\expval{\mathcal{O}_{\sigma,\zeta+\origin,\xi+\origin}}$ recursion $\green$ \eqref{eq:recursion3}, and the commutator recursion $\purple$ \eqref{eq:recursion2}.}
         \label{fig:basic_recursions}
     \end{subfigure}
     \hfill
     \begin{subfigure}[b]{0.49\textwidth}
         \centering
         \includegraphics[width=\textwidth]{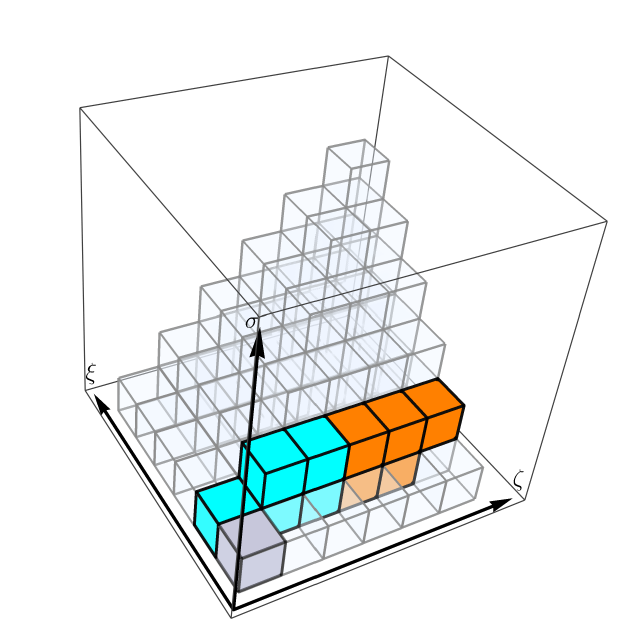}
         \caption{The derived wall recursions $\cyan$ \eqref{eq:recursion6} and $\orange$ \eqref{eq:recursion5}, obtained as linear combinations of the basic recursions. These close on the correlator-wall and are used to generate the first two floors of the cone.}
         \label{fig:modified_recursions}
     \end{subfigure}
     \caption{Diagrammatic representation of the recursion structure on the anomaly-free correlator-cone. The basic recursions do not by themselves close on the correlator-wall, but suitable linear combinations produce the wall recursions $\cyan$ and $\orange$ used in Step 1. The gray block marks the reference point $(\sigma,\zeta,\xi)=(0,0,0)$.}
     \label{fig:recursions}
\end{figure}

Note that $\purple$ \eqref{eq:recursion2} and $\green$ \eqref{eq:recursion3} cannot by themselves be placed along the correlator-wall.
However, by taking the following linear combinations,
\begin{align}
    (\orange)_{\sigma, \zeta+\origin, \xi+\origin} \coloneq & \ (\purple)_{\sigma +1,\zeta+\origin +2,\xi+\origin+1} + j_{11}(\green)_{\sigma ,\zeta +\origin+1,\xi+\origin}+ j_{12}(\green)_{\sigma ,\zeta +\origin+2,\xi+\origin},  \label{eq:construct1}\\
    (\cyan)_{\sigma, \zeta+\origin, \xi+\origin} \coloneq  & \ j_{21} (\orange)_{\sigma ,\zeta+\origin ,\xi+\origin} + j_{22} (\purple)_{\sigma ,\zeta+\origin+1 ,\xi+\origin+1} + j_{23} (\yellow)_{\sigma +1,\zeta+\origin +1,\xi+\origin} \nonumber \\ 
    &+ j_{24}(\yellow)_{\sigma +1,\zeta +\origin+2,\xi+\origin},\label{eq:construct2}
\end{align}
we obtain two height-two recursions with fixed $\xi$, where the coefficients are given in Appendix \ref{appendix:B}.
These constructions are most easily understood diagrammatically.
The new recursions read
\begin{align}
\label{eq:recursion5}
     (\orange)_{\sigma, \zeta+\origin, \xi+\origin}: 0=& \ h_{31} \mathcal{A}_{\partial \Omega}(\mathcal{O}_{\sigma+1 ,\zeta +\origin+2,\xi +\origin+1} )+h_{32} \mathcal{A}_{\partial \Omega}(\mathcal{O}_{\sigma ,\zeta +\origin+1,\xi +\origin})\nonumber \\& +h_{33} \mathcal{A}_{\partial \Omega}(\mathcal{O}_{\sigma ,\zeta +\origin+2,\xi +\origin} )+g_{31} f_{\sigma ,\zeta +\origin+1,\xi +\origin}+g_{32}f_{\sigma ,\zeta +\origin+2,\xi +\origin}\nonumber \\&+g_{33}f_{\sigma +1,\zeta +\origin+1,\xi +\origin} +g_{34} f_{\sigma +1,\zeta +\origin+2,\xi +\origin}+ g_{35}  f_{\sigma +1,\zeta +\origin+3,\xi +\origin}, 
\end{align}
\begin{align}
\label{eq:recursion6}
     (\cyan)_{\sigma, \zeta+\origin, \xi+\origin}: 0=& \ h_{41} \mathcal{A}_{\partial \Omega}(\mathcal{O}_{\sigma ,\zeta +\origin+1,\xi +\origin })+h_{42}\mathcal{A}_{\partial \Omega}(\mathcal{O}_{\sigma ,\zeta +\origin+1,\xi +\origin+1} ) \nonumber \\ 
     &+h_{43} \mathcal{A}_{\partial \Omega}(\mathcal{O}_{\sigma ,\zeta +\origin+2,\xi +\origin})+h_{44} \mathcal{A}_{\partial \Omega}(\mathcal{O}_{\sigma +1,\zeta +\origin+2,\xi +\origin+1} )\nonumber \\ 
     &+g_{41} f_{\sigma ,\zeta +\origin,\xi +\origin}  +g_{42} f_{\sigma ,\zeta +\origin+1,\xi +\origin} +g_{43}f_{\sigma ,\zeta +\origin+2,\xi +\origin}\nonumber \\ 
     & + g_{44} f_{\sigma+1 ,\zeta +\origin+1,\xi +\origin}+g_{45}f_{\sigma +1,\zeta +\origin+2,\xi +\origin}, 
\end{align}
where the coefficients are again given explicitly in Appendix \ref{appendix:B}.
Setting $\xi = 1$ and $\sigma = 0, 1$ places these two recursions along the correlator-wall, as shown in Figure \ref{fig:recursions}.

Given five unknowns, 
\begin{equation}
\{E_a, f_{0,\origin,\origin+1},f_{0,\origin+1,\origin+1}, f_{1,\origin+2, \origin+1}, f_{1,\origin+1,\origin+1}\},
\end{equation}
the recursions $\orange$ \eqref{eq:recursion5} and $\cyan$ \eqref{eq:recursion6} can be iterated to generate the entire correlator-wall.
This is illustrated in Figure \ref{fig:modified_recursions}, which shows how $\orange$ and $\cyan$ fit along the correlator-wall.

Finally, we impose the fact that the wavefunctions $\psi(z)$ are either \emph{symmetric} or \emph{antisymmetric} about $z=1/2$ \cite{Vegh:2025kgx}.
This is achieved by setting $f_{\sigma, \zeta, \xi}=0$ for $\zeta = \xi$ and odd $\sigma$, since these are integrals over $z\in[0,1]$ of odd integrands about $z = 1/2$.
Additionally, we impose 
\begin{equation}
    f_{\sigma, \zeta+\origin, \xi+\origin}= \begin{cases}
        f_{\sigma, \xi+\origin, \zeta+\origin}, \ \text{for even} \ \sigma,  \\ 
        -f_{\sigma, \xi+\origin, \zeta+\origin}, \ \text{for odd} \ \sigma .
    \end{cases}
\end{equation}
This implies $f_{1, \origin+1, \origin+1} = 0$, so we are left with only four unknowns.
Moreover, the recursion $\purple$ \eqref{eq:recursion2} and the expansion relation $\yellow$ \eqref{eq:recursion4} can be combined to solve for $f_{1, \origin+2, \origin+1}$:
\begin{align}
    0 =& \ (\purple)_{\sigma, \zeta+\origin+1, \xi+\origin+1} + (\zeta +\origin+1) (-2 \Delta +\zeta +\origin+2) (\yellow)_{\sigma ,\zeta+\origin ,\xi+\origin}\nonumber \\ 
    &-(\zeta +\xi +2 \origin+\sigma +3) (-4 \Delta +\zeta +\xi +2 \origin-\sigma +4) (\yellow)_{\sigma ,\zeta+\origin+1,\xi+\origin},
\end{align}
giving
\begin{equation}
    f_{1, \origin+2, \origin+1} = -\frac{i (\origin+1) }{2}f_{0,\origin,\origin+1}+\frac{i (2 \origin+3) }{2}f_{0,\origin+1,\origin+1}.
\end{equation}
These reduce the total number of unknowns to three:
\begin{equation}
    \{E_a, f_{0,\origin, \origin+1}, f_{0, \origin+1, \origin+1}\}.
\end{equation}

\textbf{Step 2. Generate $f_{\sigma, \zeta+\origin, \xi+\origin}$ for $\sigma = 0, 1$.}
Given the correlator-wall $f_{\sigma, \zeta+\origin, \origin+1}$ from Step 1, we can use the expansion recursion relation, $\yellow$ \eqref{eq:recursion4}, together with the fact that $\psi(z)$ is either symmetric or antisymmetric about $z=1/2$, to generate the first two floors of the anomaly-free correlator-cone.

\textbf{Step 3. Generate $f_{\sigma, \zeta+\origin, \xi+\origin}$ for $\sigma \geq 2$.}
Given the correlator-wall from Step 1 and the recursion relation $\green$ \eqref{eq:recursion3}, we can generate all higher floors of the anomaly-free correlator-cone by solving recursively for a correlator on a higher floor in terms of correlators on the lower floors.

The result is that the entire anomaly-free correlator-cone depends only on the above three unknowns, namely the energy and two correlators; the anomaly-free correlator-cone is therefore \emph{closed}.
Since the dependence of the correlators on $f_{0,\origin,\origin+1}$ and $f_{0,\origin+1,\origin+1}$ is linear, we may equivalently choose any other two correlators as the unknowns.
For symmetry between $Z$ and $1-Z$, we choose $f_{0,\origin,\origin}$ and $f_{0,\origin+1,\origin+1}$ as the two unknown correlators.

As an illustrative example, for $\Delta = 1/6$ and $\origin = 0$, we find
\begin{equation}
    f_{3,5,4}=-\frac{5 i \{7 [9 E_a (5 E_a+538)-4760] f_{0,0,0}+6 [E_a (9 E_a (13 E_a-297)-21790)+33320] f_{0,1,1}\}}{1408 (12 E_a-161) (12 E_a-85)},
\end{equation}
with poles at $E_a = 85/12$ and $E_a = 161/12$.
More generally, one can infer the candidate pole locations from Step 1.
The singularities in the correlators appear at the zeroes of $g_{35}$ in $\orange$ \eqref{eq:recursion5} and $g_{43}$ in $\cyan$ \eqref{eq:recursion6} along the correlator-wall ($\xi = 1$, $\sigma= 1, 2$), unless specific cancellations occur.
Reading off the coefficient list in Appendix \ref{appendix:B}, and noting that $g_{35}$ does not depend on $E_a$, the candidate poles are found at
\begin{equation}
    0=g_{43} \implies \{E_a\} \subseteq\left\{\frac{1}{4} (\zeta +2 \origin+3) (-8 \Delta +\zeta +2 \origin+5), \, \zeta \in \mathbb{Z}_{>1}\right\}.
\end{equation}
For the above example, with $\Delta = 1/6$ and $\origin = 0$, this gives
\begin{equation}
    \{E_a\} \subseteq \left\{\frac{85}{12}, 10, \frac{161}{12}, \frac{52}{3},\dots\right\}.
\end{equation}
In the range checked numerically, $\zeta \leq 30$, the candidate poles with odd $\zeta$ cancel. These cancellations are not used in the construction below.

\subsection{Constraint equations from the anomalous sector}
\label{sec:3.3}

The anomaly-free correlator-cone does not by itself contain any input from the boundary conditions \cite{Berenstein:2022ygg}.
To include the boundary information, we consider correlators with $\mathcal{A}_{\partial \Omega} \neq 0$.
Writing out \eqref{eq:anomaly3} explicitly:
\begin{align}
\label{eq:anomaly4}
    \mathcal{A}_{\partial \Omega}(\mathcal{O}_{\sigma, \zeta, \xi}) &= i^\sigma \biggl\{z(1-z)\left[\psi'(z)^*\partial_z^\sigma\left[z^\zeta (1-z)^\xi\psi(z)\right]-\psi(z)^*\partial_z^{\sigma+1}\left[z^\zeta(1-z)^\xi \psi(z)\right]\right] \nonumber \\
    &\qquad +(2\Delta-1)(2z-1)\psi(z)^*\partial_z^{\sigma}\left[z^\zeta(1-z)^\xi \psi(z)\right]\biggr\}\biggr\vert^{1-\epsilon}_\epsilon,
\end{align}
we see that the leading Robin boundary conditions in \eqref{eq:boundaryconditions} are not sufficient to probe the anomalous sector.
To probe $\mathcal{A}_{\partial \Omega}(\mathcal{O}_{\sigma, \zeta, \xi})$ for the parameters in the anomalous correlator-cone near $(\sigma, \zeta, \xi) = (0,\origin,\origin)$, we need the wavefunction to first subleading order in $z$ and $1-z$.
This gives 
\begin{equation}
\label{eq:boundaryconditions2}
    \psi(z) = \begin{cases}
        c_A + r c_A z^{1-2\Delta}  -\dfrac{c_A E_a}{2\Delta} z& {\rm for \ }z\rightarrow 0, \\
        \tilde c_A + r \tilde c_A(1-z)^{1-2\Delta} -\dfrac{\tilde c_A E_a}{2\Delta} (1-z) & {\rm for \ }z \rightarrow 1.
    \end{cases}
\end{equation}
Since the differential equation and the Robin parameter $r$ are real, the eigenfunctions may be chosen real up to an overall phase. 
We fix this phase by taking $c_A \in \mathbb{R}_{\neq 0}$ and $c_A = \pm\tilde{c}_A$.
The remaining overall normalization of the wavefunction is arbitrary and will be fixed below by setting $c_A = 1$.
However, we will retain $c_A$ to explicitly demonstrate the degeneracies of the boundary data.
In \eqref{eq:anomaly4}, divergences appear for $\sigma > \zeta$ or $\sigma > \xi$, and these must be removed.
The simplest way to proceed is not to cancel the divergent anomalies term by term, but rather to use the recursions to find finite constraints between the two unknown correlators $f_{0, \origin, \origin}$ and $f_{0, \origin + 1, \origin+1}$.
The motivation behind this step is that if these seed correlators are finite as $\epsilon \rightarrow 0$, then any equations involving only them should also be finite, and the divergent anomalies must cancel automatically.

In particular, we take
\begin{align}
     k_1(\orange)_{0,\origin-2,\origin} + k_2 (\purple)_{0,\origin,\origin+1} + k_3 (\cyan)_{0,\origin,\origin} + k_4(\yellow)_{1,\origin+1,\origin},
\end{align}
with the coefficients given in Appendix \ref{appendix:B}.
With this linear combination, the divergences cancel.
Then, taking $\epsilon \rightarrow 0$ and for $\origin \in \{0, 2\Delta, -1+4\Delta\}$, we find exact, independent constraints\footnote{For $\omega = 0$ and $\omega = -1+4\Delta$, the special case $\Delta = 1/4$ has $k_i =0, \, \forall i \in \{1,2,3,4\}$. However, for these two families, we can instead define $\kappa_i\coloneq k_i/(\Delta-1/4)$ and arrive at the same constraints.} given by
\begin{align}
    2  (1-2 \Delta )^2 c_A^2=& \ (4 \Delta -3)  (6 \Delta +E_a-4) f_{0,0,0} - 4 (\Delta -1) (24 \Delta +4E_a-15)f_{0,1,1}, \label{eq:constraints1}\\
    4  (1-2 \Delta )^2 \Delta(\Delta -1) c_A^2  r =& \  \left[4 \Delta  (2 \Delta +1) (4 \Delta -5) (4 \Delta -3)+3 E_a^2+4 (\Delta -1) (8 \Delta +3) E_a\right] \nonumber \\
    &\times f_{0,2 \Delta ,2 \Delta}- 4 [2 \Delta  (4 \Delta -3)+E_a] [(4 \Delta -5) (4 \Delta +3)+4 E_a] \nonumber \\
    &\times  f_{0,2 \Delta +1,2 \Delta +1}, \label{eq:constraints2}\\
    4  (1-2 \Delta)^2 \Delta ^3 c_A^2 r^2 =& \ 2 \Delta ^2 \left[24 \Delta ^2 (2 \Delta +1)  +(4 \Delta +1) E_a^2-4 \Delta  (8 \Delta +3) E_a\right] \nonumber \\& \times f_{0,-1+4 \Delta ,-1+4 \Delta}-4 (2 \Delta +1) \Delta ^2 \biggl(48 \Delta ^2+6 \Delta +4 E_a^2\nonumber\\
    &-32 \Delta  E_a-3 E_a\biggr) f_{0,4 \Delta ,4 \Delta} . \label{eq:constraints3}
\end{align}
These constraints have different dependence on the boundary conditions: $c_A^2$, $c_A^2 r$, and $c_A^2 r^2$. 
We observe that, unlike in previous QM bootstraps, the integer $\origin=0$ family does not contain all of the boundary information; the integer family is degenerate in $r$.
Similar degeneracies exist for the $\origin = 2\Delta$ and $\origin=-1+4\Delta$ families, so all three families must be used together to eliminate the degeneracy.
We discuss this point further in Section \ref{sec:5}.
For $0< \Delta < 1/2$, $\Delta = 1/4$ is a special case, where two of the three families coincide:
\begin{equation}
    0 = -1+4\Delta.
\end{equation}

We can also find separate constraints linking these three families of operators.
Let $\origin=0$ be the reference family and Taylor expand the unknown correlators of the other families, namely $(f_{0, 2\Delta, 2\Delta}$, $f_{0, 2\Delta + 1, 2\Delta +1}$, $f_{0, -1+4\Delta, -1+4\Delta}$, $f_{0, 4\Delta, 4\Delta})$, in terms of the unknown correlators of the reference family $(f_{0,0,0}, f_{0,1,1})$.
For example, for $f_{0, 2\Delta, 2\Delta}$:
\begin{align}
    f_{0,2\Delta, 2\Delta} &=\langle Z^{2\Delta}(1-Z)^{2\Delta}\rangle = \sum_{n=0}^\infty c_n \langle Z^n\rangle \\
    &= \sum_{n=0}^\infty c_n (d_{n;0} f_{0,0,0} + d_{n;1} f_{0,1,1}),
\end{align}
where $Z^{2\Delta}(1-Z)^{2\Delta}=\sum_{n=0}^\infty c_n Z^n$ is expanded about $Z=1/2$, and going from the first to second line we have used the results of the anomaly-free correlator-cone in Section \ref{sec:3.2}, writing $f_{0,n,0}$ in terms of $f_{0,0,0}$ and $f_{0,1,1}$.
We also assume that these correlators are finite, so that the integrals and sums may be exchanged.
In practice, we truncate the Taylor expansion at finite $N$.
We discuss the convergence properties in Section \ref{sec:4.1}.

With these Taylor expansions, the constraints in \eqref{eq:constraints1}--\eqref{eq:constraints3} become three equations for the three unknowns $(E_a, f_{0,0,0}, f_{0,1,1})$.
Hence, at the level of the equality constraints, the system is \emph{fully determined}; positivity is not required.

\section{Results}
\label{sec:4}

\subsection{Determining the spectrum}
\label{sec:4.1}

Before carrying out the Taylor expansions, we make a slight modification for the non-integer families.
For $\origin=2\Delta$ and $\origin=-1+4\Delta$, instead of expanding $f_{0,\origin,\origin}$ and $f_{0,\origin+1,\origin+1}$ directly, we trade them for $f_{0,\origin+2,\origin+2}$ and $f_{0,\origin+3,\origin+3}$ using the solved anomaly-free correlator-cone.
The reason is that the Taylor expansions of $[Z(1-Z)]^{\origin+2}$ and $[Z(1-Z)]^{\origin+3}$ have smaller truncation errors at finite order than those of $[Z(1-Z)]^{\origin}$ and $[Z(1-Z)]^{\origin+1}$.
Indeed, Appendix \ref{appendix:D} shows that the expansion of $[Z(1-Z)]^{b}$ converges absolutely, and that the truncation error is bounded above by a decaying power law of order $O(N^{-b})$.

Using the anomaly-free correlator-cone constructed in Section \ref{sec:3.2}, we find
\begin{align}
\label{eq:betterconvergence1}
    f_{0,\origin,\origin} &= p_2 f_{0,\origin+2,\origin+2}+p_3 f_{0,\origin+3,\origin+3}, \\
\label{eq:betterconvergence2}
    f_{0,\origin+1,\origin+1} &= q_2 f_{0,\origin+2,\origin+2}+q_3 f_{0,\origin+3,\origin+3},
\end{align}
where the coefficients are given in Appendix \ref{appendix:D}.

We now demonstrate the direct bootstrap for two representative values of $\Delta$.

\subsection{$\Delta = 1/4$}
\label{sec:4.2}

For $\Delta = 1/4$, two of the three operator families coincide, since $0=-1+4\Delta$.
Thus only the $\origin=2\Delta$ family needs to be Taylor expanded.
For $\origin=2\Delta$, \eqref{eq:betterconvergence1} and \eqref{eq:betterconvergence2} give
\begin{align}
    f_{0,\frac{1}{2},\frac{1}{2}}
    &=  \frac{4}{315} \left[\frac{7}{2} \left(16 E_a^2-464 E_a+3904\right) f_{0,\frac{5}{2},\frac{5}{2}}-32 (64-4 E_a) (26-2 E_a) f_{0,\frac{7}{2},\frac{7}{2}}\right], \\
    f_{0,\frac{3}{2},\frac{3}{2}}
    &= \frac{2}{105} \left[14 (50-2 E_a) f_{0,\frac{5}{2},\frac{5}{2}}-32 (64-4 E_a) f_{0,\frac{7}{2},\frac{7}{2}}\right].
\end{align}
Substituting these into \eqref{eq:constraints2}, we can Taylor expand the remaining fractional-power correlators in terms of $f_{0,0,0}$ and $f_{0,1,1}$.
Together with \eqref{eq:constraints1} and \eqref{eq:constraints3}, this gives three constraints for the three unknowns
\begin{equation}
    (E_a,\; f_{0,0,0},\; f_{0,1,1}).
\end{equation}

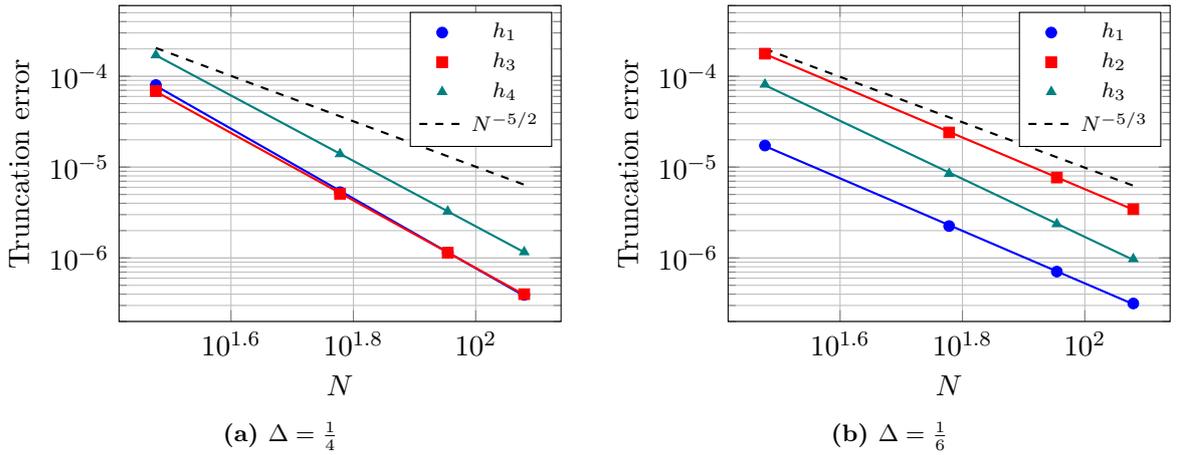
\begin{figure}[t]
\centering

\begin{subfigure}[t]{0.48\textwidth}
\centering
\begin{tikzpicture}
\begin{axis}[
  width=\textwidth,
  height=0.78\textwidth,
  xmode=log,
  ymode=log,
  xlabel={$N$},
  ylabel={Truncation error},
  ymin=2e-7,
  ymax=6e-4,
  grid=both,
  legend style={
    at={(0.98,0.98)},
    anchor=north east,
    draw=black,
    fill=white,
    font=\scriptsize
  }
]

  \addplot[only marks, mark=*,blue]
    table {plotsdat/h1_data_1_4.dat};
  \addlegendentry{$h_1$}

  \addplot[blue,thick, forget plot]
    table {plotsdat/h1_fit_1_4.dat};

  \addplot[only marks, mark=square*,red]
    table {plotsdat/h3_data_1_4.dat};
  \addlegendentry{$h_3$}

  \addplot[red,thick, forget plot]
    table {plotsdat/h3_fit_1_4.dat};

  \addplot[only marks, mark=triangle*,teal]
    table {plotsdat/h4_data_1_4.dat};
  \addlegendentry{$h_4$}

  \addplot[teal,thick, forget plot]
    table {plotsdat/h4_fit_1_4.dat};

  \addplot[black, dashed, thick]
    table {plotsdat/bound_1_4.dat};
  \addlegendentry{$N^{-5/2}$}

\end{axis}
\end{tikzpicture}
\caption{$\Delta=\tfrac14$}
\label{fig:conv14}
\end{subfigure}
\hfill
\begin{subfigure}[t]{0.48\textwidth}
\centering
\begin{tikzpicture}
\begin{axis}[
  width=\textwidth,
  height=0.78\textwidth,
  xmode=log,
  ymode=log,
  xlabel={$N$},
  ylabel={Truncation error},
  ymin=2e-7,
  ymax=6e-4,
  grid=both,
  legend style={
    at={(0.98,0.98)},
    anchor=north east,
    draw=black,
    fill=white,
    font=\scriptsize
  }
]

  \addplot[only marks, mark=*,blue]
    table {plotsdat/h1_data_1_6.dat};
  \addlegendentry{$h_1$}

  \addplot[blue,thick, forget plot]
    table {plotsdat/h1_fit_1_6.dat};

  \addplot[only marks, mark=square*,red]
    table {plotsdat/h2_data_1_6.dat};
  \addlegendentry{$h_2$}

  \addplot[red,thick, forget plot]
    table {plotsdat/h2_fit_1_6.dat};

  \addplot[only marks, mark=triangle*,teal]
    table {plotsdat/h3_data_1_6.dat};
  \addlegendentry{$h_3$}

  \addplot[teal,thick, forget plot]
    table {plotsdat/h3_fit_1_6.dat};

  \addplot[black, dashed, thick]
    table[
      x index=0,
      y expr=16*\thisrowno{1}
    ] {plotsdat/bound_1_6.dat};
  \addlegendentry{$N^{-5/3}$}

\end{axis}
\end{tikzpicture}
\caption{$\Delta=\tfrac16$}
\label{fig:conv16}
\end{subfigure}

\caption{Absolute truncation errors of selected eigenvalues as functions of the Taylor-expansion cutoff $N$. The solid lines are linear least-squares fits on a log-log scale, and the dashed lines show the reference upper-bound scaling expected from Appendix \ref{appendix:D}.}
\label{fig:convergence}
\end{figure}

Solving for $E_a$, we find good agreement with the exact spectrum.
For $N=30,60,90,120$, the first six values of $h$ are shown in Table \ref{table:delta14}, together with the exact values.
The lowest fermionic state, $h_2$, is reproduced exactly by the direct bootstrap.
In Figure \ref{fig:conv14}, we plot the absolute truncation errors of $h_1$, $h_3$, and $h_4$ as functions of $N$, together with linear least-squares fits on a log-log scale.
The fitted powers are $-3.84$, $-3.72$, and $-3.60$, respectively.
These are faster than the upper bound $O(N^{-5/2})$ coming from the expansion of $f_{0,5/2,5/2}$.

\begin{figure}
    \centering
    \includegraphics[width=0.85\linewidth]{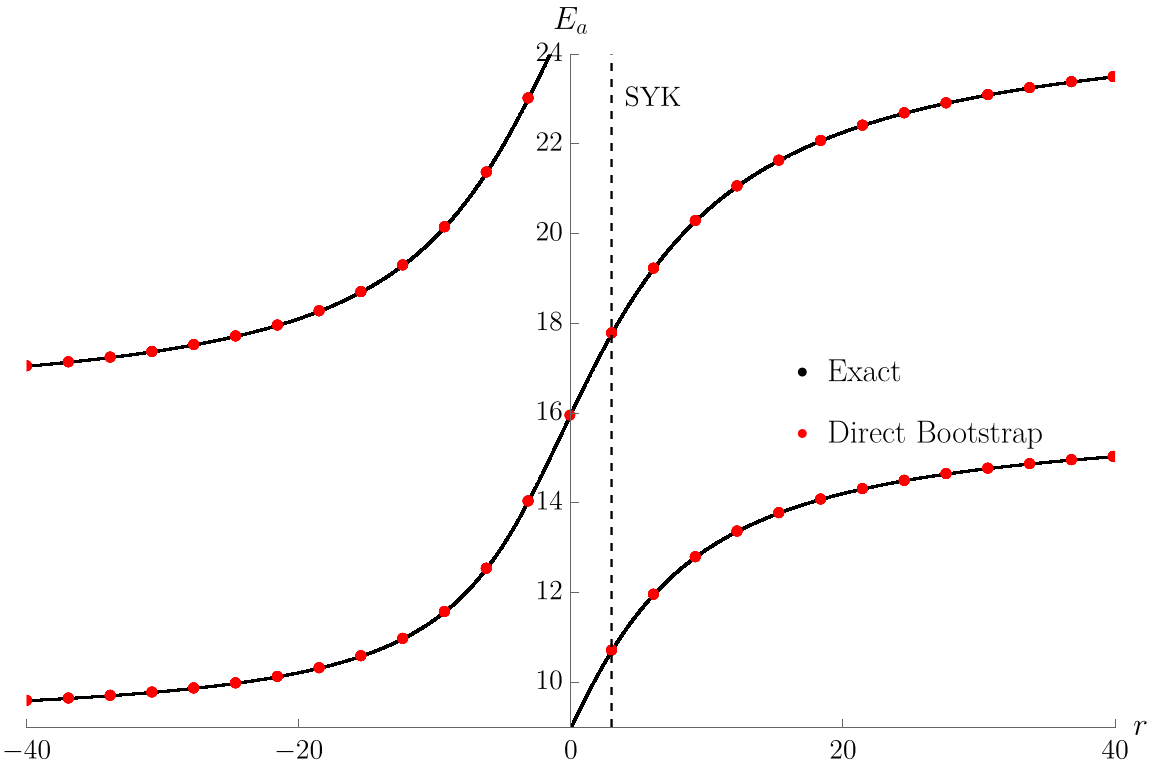}
    \caption{$\Delta=\frac14$ at truncation order $N=120$. Energy eigenvalues for general Robin boundary conditions parametrized by $r$. The red dots are the direct-bootstrap results, the black curves are the exact spectrum from the analytic wavefunctions, and the dotted vertical line marks the value of $r$ that reproduces the SYK spectrum.}
    \label{fig:generalr2}
\end{figure}

We also test the method away from the SYK value of the Robin parameter by allowing general boundary conditions, that is, arbitrary $r$.
The result is shown in Figure \ref{fig:generalr2}.
Again we find good agreement with the exact spectrum.
Interestingly, even for boundary conditions without fractional powers ($r=0$), the fractional-power operator families still provide the additional constraints \eqref{eq:constraints2} and \eqref{eq:constraints3}, allowing the spectrum to be determined without imposing positivity.

\begin{table}
\centering
\begin{tabular}{l*{6}{r}}
\toprule
 & $h_{1}$ & $h_{2}$ & $h_{3}$ & $h_{4}$ & $h_{5}$ & $h_{6}$ \\
\midrule
$N=30$  & $\underline{1.216}2775$ & $\underline{2.0000000}$ & $\underline{2.860}4625$ & $\underline{3.773}3648$ & $\underline{4.71}83591$ & $\underline{5.6}984770$ \\
$N=60$  & $\underline{1.21635}23$ & $\underline{2.0000000}$ & $\underline{2.86039}88$ & $\underline{3.7735}217$ & $\underline{4.717}6114$ & $\underline{5.6}812811$ \\
$N=90$  & $\underline{1.21635}64$ & $\underline{2.0000000}$ & $\underline{2.86039}49$ & $\underline{3.77353}24$ & $\underline{4.7175}551$ & $\underline{5.679}9255$ \\
$N=120$ & $\underline{1.216357}2$ & $\underline{2.0000000}$ & $\underline{2.86039}41$ & $\underline{3.77353}45$ & $\underline{4.7175}436$ & $\underline{5.679}6346$ \\
\midrule
$\mathrm{exact}$ & $1.2163575$ & $2.0000000$ & $2.8603937$ & $3.7735356$ & $4.7175370$ & $5.6794590$ \\
\bottomrule
\end{tabular}
\caption{$\Delta=\frac14$ with $r=\frac{1-\Delta}{\Delta}$. First six eigenvalues obtained from the direct bootstrap at different truncation orders $N$, compared with the exact values. Underlining marks the leading digits that agree with the exact value. The state $h_2=2$ is reproduced exactly.}
\label{table:delta14}
\end{table}

\FloatBarrier

\subsection{$\Delta = 1/6$}
\label{sec:4.3}

\begin{table}
\centering
\begin{tabular}{l*{6}{r}}
\toprule
 & $h_{1}$ & $h_{2}$ & $h_{3}$ & $h_{4}$ & $h_{5}$ & $h_{6}$ \\
\midrule
$N=30$ & $\underline{1.3623}073$ & $\underline{2.000}1773$ & $\underline{2.728}2777$ & $\underline{3.5776}445$ & $\underline{4.49}76231$ & $\underline{5.45}25264$ \\
$N=60$ & $\underline{1.36232}23$ & $\underline{2.0000}241$ & $\underline{2.728}2049$ & $\underline{3.57763}72$ & $\underline{4.498}5731$ & $\underline{5.45}35061$ \\
$N=90$ & $\underline{1.36232}39$ & $\underline{2.00000}77$ & $\underline{2.72819}88$ & $\underline{3.57763}40$ & $\underline{4.4987}086$ & $\underline{5.45}39178$ \\
$N=120$ & $\underline{1.362324}3$ & $\underline{2.00000}34$ & $\underline{2.72819}74$ & $\underline{3.57763}29$ & $\underline{4.4987}487$ & $\underline{5.454}0782$ \\
\midrule
$\mathrm{exact}$ & $1.3623246$ & $2.0000000$ & $2.7281964$ & $3.5776317$ & $4.4987859$ & $5.4542702$ \\
\bottomrule
\end{tabular}
\caption{$\Delta=\frac16$ with $r=\frac{1-\Delta}{\Delta}$. First six eigenvalues obtained from the direct bootstrap at different truncation orders $N$, compared with the exact values. Underlining marks the leading digits that agree with the exact value.}
\label{table:delta16}
\end{table}

For $\origin=2\Delta$, \eqref{eq:betterconvergence1} and \eqref{eq:betterconvergence2} give
\begin{align}
    f_{0,\frac13,\frac13}
    &= \frac{7 [9 E_a (45 E_a-1294)+96248] f_{0,\frac73,\frac73}-40 (9 E_a-115) (36 E_a-575) f_{0,\frac{10}{3},\frac{10}{3}}}{3696},\\
    f_{0,\frac43,\frac43}
    &= -\frac{2\left[7 (9 E_a-223) f_{0,\frac73,\frac73}+8 (575-36 E_a) f_{0,\frac{10}{3},\frac{10}{3}}\right]}{231}.
\end{align}
For $\origin=-1+4\Delta$, we have
\begin{align}
    f_{0,-\frac13,-\frac13}
    &= \frac{1}{140} [E_a (561 E_a-10150)+48216] f_{0,\frac53,\frac53}-\frac{11}{7} (E_a-7) (12 E_a-133) f_{0,\frac83,\frac83}, \\
    f_{0,\frac23,\frac23}
    &= \frac{1}{105} \left[34 (49-3 E_a) f_{0,\frac53,\frac53}+40 (12 E_a-133) f_{0,\frac83,\frac83}\right].
\end{align}
Substituting these into \eqref{eq:constraints1}--\eqref{eq:constraints3}, we can again Taylor expand the fractional-power correlators in terms of $f_{0,0,0}$ and $f_{0,1,1}$.
This gives three constraints for three unknowns.

Solving for $E_a$, we again find good agreement with the exact spectrum.
For $N=30,60,90,120$, the first six values of $h$ are shown in Table \ref{table:delta16}, together with the exact values.
In Figure \ref{fig:conv16}, we plot the absolute truncation errors of $h_1$, $h_2$, and $h_3$ as functions of $N$, together with linear least-squares fits on a log-log scale.
The fitted powers are $-2.89$, $-2.85$, and $-3.19$, respectively. 
These are faster than the upper bound $O(N^{-5/3})$ coming from the expansion of $f_{0,5/3,5/3}$.

\clearpage
\section{Discussion}
\label{sec:5}

In this paper, we applied the quantum mechanical bootstrap to a system on the closed interval $0 \le z \le 1$ whose spectrum coincides with the bilinear operator spectrum of the Sachdev--Ye--Kitaev model.
The Schr\"odinger equation is exactly solvable, making it possible to compare the bootstrap results against the exact eigenvalues at every stage.
The wavefunctions obey fractional-power Robin boundary conditions parametrized by $r$, and the self-adjoint extension is specified by matching to the SYK conformal dimension $\Delta$.
Taking expectation values of commutators with the operators
\begin{equation}
    \mathcal{O}_{\sigma,\zeta,\xi} = S^\sigma Z^\zeta (1-Z)^\xi
\end{equation}
yields recursion relations for the correlators $f_{\sigma,\zeta,\xi} := \expval{\mathcal{O}_{\sigma,\zeta,\xi}}$, together with dressed anomalies $\mathcal{A}(\mathcal{O}_{\sigma,\zeta,\xi})$ that arise whenever $w^{-1}\mathcal{O}_{\sigma,\zeta,\xi}$ fails to preserve the domain of $H_a$.
In the dressed correlator variables, the commutator identity splits into a boundary/domain anomaly and a bulk correction.  
The boundary term carries the fractional Robin data, while the bulk correction extends the recursion relations but contains no independent boundary information.

Since the recursion relations preserve the fractional parts of $\zeta$ and $\xi$, the operators decompose into families labelled by a common fractional offset $\origin$:
\begin{equation}
    \mathcal{O}_{\sigma,\,\origin+\zeta,\,\origin+\xi}, \qquad \sigma,\zeta,\xi \in \mathbb{Z}_{\ge 0}.
\end{equation}
Within each family, the anomaly-free correlator-cone closes on three unknowns---the energy and two correlators---but is insensitive to the boundary conditions.
To probe the boundary data, one must consider operators with finite, nonzero boundary anomalies.
These give exact constraint equations for three special families,
\begin{equation}
    \origin \in \{0,\, 2\Delta,\, 4\Delta - 1\},
\end{equation}
each of which depends on a different power of the boundary parameter $r$.
 
An important observation is that without the cross-family constraints, the positivity generated by these operator families---the standard tool of the QM bootstrap---is not sufficient to determine the spectrum of this model.
The reason is that the anomalous constraints contain only degenerate information about the boundary parameter $r$. For $\origin=0$, the constraint is independent of $r$. For $\origin=2\Delta$, $r$ appears only through $c_A^2 r$, so it is degenerate with the normalization constant $c_A$. For $\origin=4\Delta-1$, it appears through $c_A^2 r^2$, which introduces an additional degeneracy by erasing the sign of $r$. Hence, even taken together, these three families do not supply enough information for positivity to fix the boundary conditions and thereby determine the spectrum. This is consistent with the discussion in Section \ref{sec:3.2} and with Figure \ref{fig:generalr2}, where $E_a(r)$ can take essentially any positive value. We also demonstrate the degeneracy in positivity bounds explicitly for the operator family at $\Delta = 1/4$ with $\origin = 2\Delta$ in Appendix \ref{appendix:C}.

What makes the problem solvable is the Taylor expansion relating the three special operator families.
This provides the cross-family input that positivity alone cannot capture.
Combined with the anomalous constraints \eqref{eq:constraints1}--\eqref{eq:constraints3}, it gives three independent equations for three unknowns $(E_a, f_{0,0,0}, f_{0,1,1})$, allowing the spectrum to be determined directly without invoking positive semidefiniteness.
In this sense, the Taylor expansion is not just a technical device; it is the structural ingredient that lifts the degeneracies left by the anomalous sector.
This is why we refer to the method as the \emph{direct bootstrap}: the spectrum follows from \emph{equations}, not from \emph{bounds}.

Surprisingly, even for integer-powered boundary conditions ($r=0$), the exact constraints from the fractional-power operator families, $\origin \in \{2\Delta, 4\Delta - 1\}$, remain valid and provide the additional equations needed to fix the spectrum, as demonstrated in Figure~\ref{fig:generalr2}.

More generally, this model illustrates both the power and the limitations of the QM bootstrap in the presence of fractional boundary conditions.
Domain anomalies provide a controlled mechanism by which boundary data enters the bootstrap recursion relations, but anomalous constraints need not always suffice on their own.
In the present case, the spectrum becomes accessible only because the Taylor expansion supplies additional non-degenerate relations across operator families.

In more general QM models with fractional Robin boundary conditions, where the number of unknowns may exceed the number of available exact relations, positivity is expected to provide the additional input needed to close the system. The present model therefore gives a concrete example in which positivity alone is not sufficient, and where the spectrum becomes accessible only after incorporating exact relations across operator families.

Several natural extensions of this work suggest themselves: the direct bootstrap could be applied to \emph{other models with general boundary conditions} (e.g. hypergeometric or Heun's equations).
These would serve as further testing grounds for the interplay between anomalous constraints and cross-family Taylor expansions, and could help clarify the conditions under which the direct bootstrap closes. For more complicated Hamiltonians it remains to be seen whether analogous cross-family relations exist and whether they provide sufficient constraints.
A hybrid approach, combining a finite number of exact equations from anomalous sectors with positivity from the PSD bootstrap, may prove effective in such cases.
Finally, one may ask whether the ``from bounds to equations'' mechanism identified here has analogues in higher-dimensional bootstrap programmes.

{ \bf Acknowledgements}: KHT is supported by a Science and Technology Facilities Council (STFC) studentship. The work of DV is partially supported by the STFC Consolidated Grant ST/X00063X/1 “Amplitudes, Strings, \& Duality.”
KHT thanks the organizers of the ``Spring School on Superstring Theory and Related Topics" at the ICTP, where the results were presented.
No new data were generated or analyzed during this study.

\newpage
\clearpage

\appendix

\section{Calculating the dressed anomaly}
\label{appendix:A}

For the operators and correlators
\begin{equation}
    \mathcal{O}_{\sigma,\zeta,\xi}\coloneq S^\sigma Z^\zeta (1-Z)^\xi,
    \qquad
    f_{\sigma,\zeta,\xi}\coloneq \expval{O_{\sigma, \zeta, \xi}},
\end{equation}
we compute the dressed anomaly
\begin{align}
    \mathcal{A}(\mathcal{O}_{\sigma,\zeta,\xi})
    &= \bra{\psi}\ket{H_a \mathcal{O}_{\sigma,\zeta,\xi} \psi}-\bra{H_a \psi} \ket{\mathcal{O}_{\sigma,\zeta,\xi} \psi} \nonumber \\
    &= -\int_\epsilon^{1-\epsilon} dz \,
    \left\{
    [H_a\psi(z)]^*[S^\sigma z^\zeta(1-z)^\xi \psi(z)]
    -
    \psi(z)^* [H_aS^\sigma z^\zeta (1-z)^\xi\psi(z)]
    \right\} \nonumber \\
    &= -i^\sigma \int_\epsilon^{1-\epsilon} dz \,
    \biggl\{
    [2\Delta(2z-1)\psi'(z)+z(z-1)\psi''(z)]^*
    \,\partial^{\sigma}_z\!\bigl[z^\zeta(1-z)^\xi\psi(z)\bigr]
    \nonumber \\
    &\qquad\qquad\qquad
    -\psi(z)^*
    \Bigl[
    2\Delta(2z-1)\partial^{\sigma+1}_z\!\bigl(z^\zeta(1-z)^\xi \psi(z)\bigr)
    \nonumber \\&\qquad\qquad\qquad+ z(z-1)\partial^{\sigma+2}_z\!\bigl(z^\zeta (1-z)^\xi\psi(z)\bigr)
    \Bigr]
    \biggr\}.
\end{align}
Defining
\begin{equation}
    \alpha(z)\coloneq \partial^{\sigma}_z\!\bigl[z^\zeta(1-z)^\xi\psi(z)\bigr],
    \qquad
    \beta(z)\coloneq \psi(z)^*,
    \qquad
    \gamma(z)\coloneq z(1-z),
\end{equation}
the anomaly becomes
\begin{align}
    \mathcal{A}(\mathcal{O}_{\sigma,\zeta,\xi})
    &= i^\sigma \int_\epsilon^{1-\epsilon} dz \,
    \left[
    2\Delta \gamma' (\beta' \alpha-\beta \alpha')
    + \gamma (\beta'' \alpha - \beta \alpha'')
    \right] \nonumber \\
    &= i^\sigma \left[\gamma (\beta' \alpha-\beta \alpha') \right]_\epsilon^{1-\epsilon}
    + i^\sigma(2\Delta-1)\int_\epsilon^{1-\epsilon} dz \,
    \gamma' (\beta' \alpha-\beta \alpha') \nonumber \\
    &= i^\sigma \left[\gamma (\beta' \alpha-\beta \alpha') + (2\Delta-1)\gamma' \beta \alpha \right]_\epsilon^{1-\epsilon}
    - i^\sigma(2\Delta-1)\int_\epsilon^{1-\epsilon} dz \,
    \left( \gamma'' \beta \alpha+ 2\gamma' \beta \alpha' \right).
\end{align}
The first term is the boundary contribution.
The remaining bulk integral can be rewritten using repeated canonical commutators, equivalently the McCoy formulas, to obtain
\begin{align}
\label{eq:anomaly}
    \mathcal{A}(\mathcal{O}_{\sigma,\zeta,\xi})
    &= i^\sigma \left[\gamma (\beta' \alpha-\beta\alpha') + (2\Delta-1) \gamma' \beta\alpha \right]_\epsilon^{1-\epsilon}
    -2(2\sigma+1)(2\Delta-1) f_{\sigma,\zeta,\xi}\nonumber\\
    &\quad
    +2i(2\Delta-1)f_{\sigma+1,\zeta,\xi}
    -4i (2\Delta-1)f_{\sigma+1,\zeta+1,\xi}.
\end{align}
Thus, the dressed anomaly splits naturally into a boundary term and a bulk term.  
The boundary term is the dressed domain anomaly and encodes the fractional Robin data; the bulk term is the correction induced by writing the recursion relations in dressed correlators.

\section{Recursion coefficients}
\label{appendix:B}

The recursion coefficients are given by
\begin{align}
    g_{11} \coloneq& \ 2 \Delta  (3 \zeta +\xi +4 \origin+2 \sigma +2)-2 (\zeta +\origin) (\xi +\origin+2)-2 (\zeta +\origin)^2-2 (\xi +\origin)+\sigma ^2-\sigma -2,\\
    g_{12} \coloneq& \ (\zeta +\origin) (-2 \Delta +\zeta +\origin+1), \\
    g_{13} \coloneq& \ i [2 (-2 \Delta +\zeta +\origin+1)-\sigma ], \\
    g_{14} \coloneq& \ (\zeta +\xi +2 \origin+\sigma +1) (-4 \Delta +\zeta +\xi +2 \origin-\sigma +2), \\
    g_{15} \coloneq&  -2 i (-4 \Delta +\zeta +\xi +2 \origin-\sigma +2), \\
    g_{21} \coloneq&  -[E_a-\sigma  (4 \Delta +\sigma -1)] ,\\
    g_{22} \coloneq&  -i (2 \Delta +\sigma ) , \\
    g_{23} \coloneq& \ 2i (2 \Delta +\sigma ), \\
    g_{24} \coloneq& \ 1, \\
    j_{11} \coloneq& \ i (4 \Delta -2 \zeta -2\origin+\sigma -5), \\
    j_{12} \coloneq& \ 2 i (-4 \Delta +\zeta +\xi +2 \origin-\sigma +4), \\
    j_{21} \coloneq& -2i, \\
    j_{22} \coloneq& -(8 \Delta -\zeta -\xi -2\origin +3 \sigma -5), \\
    j_{23} \coloneq& \ i(8 \Delta -\zeta -\xi -2\origin +3 \sigma -5)(4 \Delta -2 \zeta -2\origin+\sigma -4), \\
    j_{24} \coloneq& \ 2i(8 \Delta -\zeta -\xi -2\origin +3 \sigma -5)(-4 \Delta +\zeta +\xi +2\origin-\sigma +4), \\
    g_{31} \coloneq&  -i [E_a-\sigma  (4 \Delta +\sigma -1)] [4 \Delta -2 (\zeta +\origin )+\sigma -5], \\
    g_{32} \coloneq& \ 2 i [E_a-\sigma  (4 \Delta +\sigma -1)][4 \Delta -\zeta -\xi +\sigma -2 \origin -4], \\
    g_{33} \coloneq& \ (-4 \Delta +\zeta -\sigma +\origin +3) (-2 \Delta +\zeta -\sigma +\origin +2), \\
    g_{34} \coloneq&  -32 \Delta ^2+2 \Delta  (9 \zeta +3 \xi -10 \sigma +12 \origin +29)-2 \zeta ^2-2 \zeta  (\xi -3 \sigma +3 \origin +7)-2 \origin  (\xi -4 \sigma +10) \nonumber \\& +2 \xi  \sigma -6 \xi -3 \sigma ^2+19 \sigma -4 \origin ^2-24, \\
    g_{35} \coloneq&  (4 \Delta -\zeta -\xi +\sigma -2 \origin -4) (8 \Delta -\zeta -\xi +3 \sigma -2 \origin -5), \\
    g_{41} \coloneq& - (\zeta +\origin +1) (-2 \Delta +\zeta +\origin +2) (8 \Delta -\zeta -\xi +3 \sigma -2 \origin -5) , \\ 
    g_{42} \coloneq& -2 [E_a-\sigma  (4 \Delta +\sigma -1)] [4 \Delta -2 (\zeta +\origin )+\sigma -5]   - (8 \Delta -\zeta -\xi +3 \sigma -2 \origin -5) \nonumber \\ & \times\biggl[2 \Delta  (3 \zeta +\xi +2 \sigma +4 \origin +6)-2 (\zeta +\origin +1) (\xi +\origin +3) -2 (\zeta +\origin +1)^2\nonumber \\&-2 (\xi +\origin +1)+\sigma ^2-\sigma -2\biggr],  \\
    g_{43} \coloneq& - (-4 \Delta +\zeta +\xi -\sigma +2 \origin +4) [(\zeta +\xi -\sigma +2 \origin +3) (8 \Delta -\zeta -\xi +\sigma -2 \origin -5)+4 E_a], \\ 
    g_{44} \coloneq& \ i  \biggl[16 \Delta ^2-4 \Delta  (2 \zeta +\xi -2 \sigma +3 \origin +6)+4 \xi +\sigma ^2-7 \sigma +2 \origin ^2 +2 \origin  (\zeta +\xi -2 \sigma +4)\nonumber \\&+2 \zeta  \xi -3 \zeta  \sigma +4 \zeta -\xi  \sigma +8\biggr], \\ 
    g_{45} \coloneq&  -i (8 \Delta -2 \xi +3 \sigma -2 \origin -4) (4 \Delta -\zeta -\xi +\sigma -2 \origin -3), \\
    h_{31} \coloneq& -1, \\
    h_{32} \coloneq&  -i (4 \Delta -2 \zeta +\sigma -2 \origin -5), \\
    h_{33} \coloneq& \ 2 i (4 \Delta -\zeta -\xi +\sigma -2 \origin -4), \\
    h_{41} \coloneq& - 2 (4 \Delta -2 (\zeta +\origin )+\sigma -5) , \\
    h_{42} \coloneq& \ (8 \Delta -\zeta -\xi +3 \sigma -2 \origin -5) , \\
    h_{43} \coloneq& - 4 (-4 \Delta +\zeta +\xi -\sigma +2 \origin +4), \\
    h_{44} \coloneq& \ 2 i ,\\
    k_1 \coloneq& \ i \origin  (-4 \Delta +\origin +2) (-2 \Delta +\origin +1),\\
    k_2 \coloneq& -E_a (4 \Delta -2 \origin -1) (-4 \Delta +\origin +2),\\
    k_3 \coloneq& \ \origin  (2 \Delta -\origin -1) (8 \Delta -2 \origin -3)+E_a (4 \Delta -2 \origin -1),\\
    k_4 \coloneq& -2 i (4 \Delta -2 \origin -3) (-4 \Delta +\origin +2) [\origin  (2 \Delta -\origin -1) (8 \Delta -2 \origin -3) \nonumber \\&+E_a (4 \Delta -2 \origin -1)].
\end{align}

\section{Degeneracies in positivity bounds}
\label{appendix:C}

In this appendix, we demonstrate explicitly the degeneracy in the positivity constraints for the operator family at $\Delta = 1/4$ with $\origin = 2\Delta$.
Using constraint \eqref{eq:constraints2}, this family depends on
\begin{equation}
    \{E_a,\; f_{0,3/2,3/2},\; c_A^2 r\}.
\end{equation}
The factor $c_A^2 r$ is precisely what leads to the degeneracy.
The ratio $r$ is fixed by \eqref{eq:boundary} but $c_A \in \mathbb{R}_{\neq 0}$. 
In what follows, we set $c_A = 1$ as an example and show that the allowed region still remains extended rather than collapsing to isolated eigenvalues.

We choose the bootstrap operator to be
\begin{equation}
    \mathcal{O}
    =
    \sum_{\substack{\sigma \geq 0 \\ \zeta \geq \sigma,\ \xi \geq \sigma}}
    c_{\sigma,\zeta,\xi}\,(1-Z)^{\xi+\origin/2} Z^{\zeta+\origin/2} S^\sigma.
\end{equation}
This choice is inspired by \cite{Sword:2024gvv}; it is the ordering that gave the tightest positivity bounds in the interval bootstrap.

Positivity, 
\begin{equation}
\label{eq:positivity}
    \bra{\mathcal{O}\psi}\ket{\mathcal{O}\psi} = \int_0^1 dz\,|\mathcal{O}\psi|^2 \geq 0,
\end{equation}
implies
\begin{equation}
    \mathcal{B}_{\rm 3d} \succeq 0,
\end{equation}
where the matrix elements are
\begin{align}
    (\mathcal{B}_{\rm 3d})_{(\sigma,\zeta,\xi),(\sigma',\zeta',\xi')}
    &\coloneq
    \bra{\bigl((1-Z)^{\xi'+\origin/2} Z^{\zeta'+\origin/2} S^{\sigma'}\bigr) \psi} \ket{(1-Z)^{\xi+\origin/2} Z^{\zeta+\origin/2} S^\sigma \psi} \nonumber \\
    &=
    \expval{S^{\sigma'} Z^{\zeta+\zeta'+\origin}(1-Z)^{\xi+\xi'+\origin} S^\sigma},
\end{align}
where the second line uses the dagger lemma of \cite{Sword:2024gvv}. 
Using the McCoy commutation formulas \cite{Mccoy:1929com}, this becomes
\begin{align}
    (\mathcal{B}_{\rm 3d})_{(\sigma,\zeta,\xi),(\sigma',\zeta',\xi')}
    =
    \sum_{\substack{0 \leq k \leq s \\ 0 \leq s \leq \sigma}}
    c(\sigma,\zeta,\xi,\zeta',\xi';s,k)\,
    \expval{S^{\sigma+\sigma'-s}Z^{\zeta+\zeta'+\origin-s+k}(1-Z)^{\xi+\xi'+\origin-k}},
\end{align}
with
\begin{equation}
    c(\sigma,\zeta,\xi,\zeta',\xi';s,k)
    \coloneq
    (-i)^s(-1)^k\,
    \frac{\Gamma(\sigma+1)}{\Gamma(\sigma-s+1)}
    \binom{\xi+\xi'+\origin}{k}\binom{\zeta+\zeta'+\origin}{s-k}.
\end{equation}

\begin{figure}
    \centering
    \includegraphics[width=0.85\linewidth]{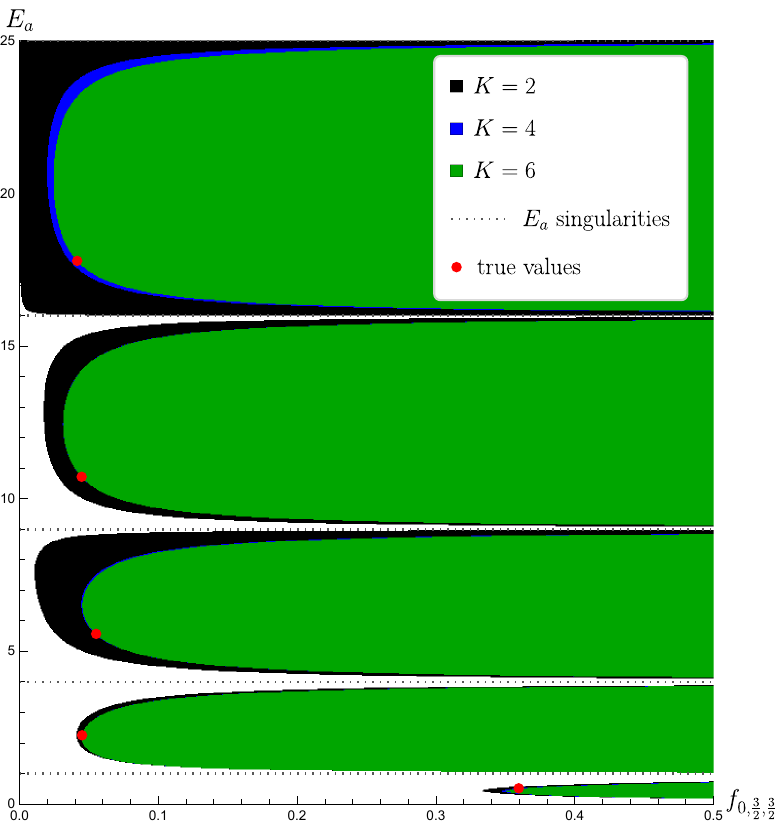}
    \caption{Bootstrap allowed regions in the $(f_{0,\frac{3}{2},\frac{3}{2}},E_a)$ plane for truncation orders $K=2,4,6$, with $\Delta=\tfrac14$, $\origin=2\Delta$, and $c_A=1$. As $K$ increases, the boundaries of the allowed regions appear to converge toward limiting curves that intersect the exact spectrum (red dots). The horizontal dotted lines denote the candidate $E_a$ singularities.}
    \label{fig:bootstrapPSD}
\end{figure}

Following \cite{Sword:2024gvv}, we then construct a $K\times K$ matrix $\mathcal{B}$ with entries
\begin{equation}
    \mathcal{B}_{i,j}
    =
    (\mathcal{B}_{\rm 3d})_{(i,2i,2i),(j,2j,2j)}.
\end{equation}
Using the solved anomaly-free cone from Section \ref{sec:3.2} together with $\mathcal{B}\succeq 0$, we find that the allowed values of $(E_a,f_{0,3/2,3/2})$ lie in large islands separated by singularities in $E_a$, as shown in Figure \ref{fig:bootstrapPSD}.
In particular, positivity alone does \emph{not} isolate the spectrum in this family.
As $K$ increases, the boundaries of the allowed regions appear to converge toward limiting curves that pass through the exact values (red dots).

\section{Convergence}
\label{appendix:D}

Consider the Taylor expansion of $[Z(1-Z)]^b$ about $Z=\frac12$, with $b\in \mathbb{R}_{>0}$.
Defining
\begin{equation}
    u\coloneq 2Z-1,
\end{equation}
we have
\begin{equation}
    [Z(1-Z)]^b
    =
    4^{-b}(1-u^2)^b
    =
    4^{-b}\sum_{n=0}^\infty \binom{b}{n}(-1)^n u^{2n}.
\end{equation}
For integer $b\ge 0$ the series truncates exactly.
For non-integer $b>0$, the generalized binomial coefficients satisfy
\begin{equation}
    \left|\binom{b}{n}\right|
    \sim
    \frac{1}{|\Gamma(-b)|}\,n^{-b-1}
    \qquad (n\to\infty),
\end{equation}
so the series converges absolutely even at the endpoints $|u|=1$.

Expanding in powers of $Z$ gives the coefficients used in Section \ref{sec:4.1}:
\begin{align}
    [Z(1-Z)]^b
    &= 4^{-b}\sum_{n=0}^\infty \binom{b}{n} (-1)^n \sum_{k=0}^{2n} \binom{2n}{k}2^k(-1)^{2n-k} Z^{k} \\ 
    &= \lim_{N\rightarrow \infty} 4^{-b}\sum_{n=0}^{N/2}\binom{b}{n} (-1)^n \sum_{k=0}^{2n} \binom{2n}{k}2^k(-1)^{2n-k} Z^{k},
\end{align}
where the second line makes explicit the truncation at the $N$th power in $Z$.
For convenience, we take $N$ to be even.

\emph{Rate of convergence}. The truncation error is
\begin{equation}
    R_N(Z)
    =
    4^{-b}\sum_{n=\frac{N}{2}+1}^\infty \binom{b}{n} (-1)^n (2Z-1)^{2n}.
\end{equation}
Since $|2Z-1|\le 1$, we obtain
\begin{equation}
    |R_N(Z)|
    \le
    4^{-b}\sum_{n=\frac{N}{2}+1}^\infty \left|\binom{b}{n}\right|
    \le
    M\sum_{n=\frac{N}{2}+1}^\infty n^{-b-1},
\end{equation}
for some positive constant $M$.
Using
\begin{equation}
    n^{-b-1}\le \int_{n-1}^{n} dx\,x^{-b-1},
\end{equation}
we find
\begin{equation}
    |R_N(Z)|
    \le
    M\int_{N/2}^{\infty} dx\,x^{-b-1}
    =
    \frac{M}{b}\left(\frac{N}{2}\right)^{-b}.
\end{equation}
Thus the truncation error is bounded above by a decaying power law of order $O(N^{-b})$.
Away from the endpoints, the actual decay can be faster because of the extra factor $(2Z-1)^{2n}$.

To improve convergence, we therefore Taylor expand correlators with larger effective exponent $b$ rather than expanding directly at the lowest member of a fractional family.
Using the anomaly-free correlator-cone from Section \ref{sec:3.2}, we find
\begin{align}
\label{eq:appendixbetter1}
    f_{0,\origin,\origin} &= p_2f_{0,\origin+2,\origin+2}+p_3f_{0,\origin+3,\origin+3}, \\
\label{eq:appendixbetter2}
    f_{0,\origin+1,\origin+1} &= q_2f_{0,\origin+2,\origin+2}+q_3f_{0,\origin+3,\origin+3},
\end{align}
where
\begin{align}
    p_2 \coloneq& \ \frac{4}{\mu} \biggl\{E_a^2 (-4 \Delta +2 \origin+5) (-4 \Delta +2 \origin+7)+4 E_a\biggl[16 \Delta ^3 (3 \origin+7)-4 \Delta ^2 (3 \origin+8) (5 \origin+14) \nonumber \\ &+\Delta  (8 \origin (\origin (3 \origin+26)+75)+579)-3 (\origin+3)^4\biggr]-12 \Delta  (\origin+3) \biggl[\origin (2 \origin (6 \origin^2+69 \origin+292)\nonumber \\&+1075)+724\biggr] +192 \Delta ^4 (2 \origin (2 \origin+9)+19)-16 \Delta ^3 (\origin (18 \origin (4 \origin+31)+1405)+1134) \nonumber \\&+4 \Delta ^2 (\origin (\origin (12 \origin (13 \origin+144)+7069)+12618)+8247)\nonumber \\&+3 (\origin+2) (\origin+3)^2 (\origin+4) (2 \origin+5) (2 \origin+7)\biggr\}, \\
    p_3 \coloneq& \ -\frac{8}{\mu}  (-4 \Delta +2 \origin+5) (-2 \Delta +\origin+4) \bigl[(2 \origin+7) (-8 \Delta +2 \origin+9)-4 E_a\bigr] \nonumber \\& \times \bigl[2 (-7 \Delta +\origin (-4 \Delta +\origin+5)+6)-E_a\bigr], \\
    \mu = & \ (\origin+1) (\origin+2) (-4 \Delta +\origin+3) (-4 \Delta +\origin+4) (-2 \Delta +\origin+2) (-2 \Delta +\origin+3), \\
    q_2 \coloneq& \ \frac{-2 (-4 \Delta +2 \origin+7) (22 \Delta +E_a-2 \origin (-4 \Delta +\origin+7)-24)}{(\origin+2) (-4 \Delta +\origin+4) (-2 \Delta +\origin+3)}, \\
    q_3 \coloneq& \ \frac{4 (-2 \Delta +\origin+4) (4 E_a-(2 \origin+7) (-8 \Delta +2 \origin+9))}{(\origin+2) (-4 \Delta +\origin+4) (-2 \Delta +\origin+3)}.
\end{align}


\bibliographystyle{JHEP}
\bibliography{paper}

@article{Maldacena:2015waa,
    author = "Maldacena, Juan and Shenker, Stephen H. and Stanford, Douglas",
    title = "{A bound on chaos}",
    eprint = "1503.01409",
    archivePrefix = "arXiv",
    primaryClass = "hep-th",
    doi = "10.1007/JHEP08(2016)106",
    journal = "JHEP",
    volume = "08",
    pages = "106",
    year = "2016"
}

@article{Kitaev:2017awl,
    author = "Kitaev, Alexei and Suh, S. Josephine",
    title = "{The soft mode in the Sachdev-Ye-Kitaev model and its gravity dual}",
    eprint = "1711.08467",
    archivePrefix = "arXiv",
    primaryClass = "hep-th",
    doi = "10.1007/JHEP05(2018)183",
    journal = "JHEP",
    volume = "05",
    pages = "183",
    year = "2018"
}

@article{Rattazzi:2008pe,
    author = "Rattazzi, Riccardo and Rychkov, Vyacheslav S. and Tonni, Erik and Vichi, Alessandro",
    title = "{Bounding scalar operator dimensions in 4D CFT}",
    eprint = "0807.0004",
    archivePrefix = "arXiv",
    primaryClass = "hep-th",
    doi = "10.1088/1126-6708/2008/12/031",
    journal = "JHEP",
    volume = "12",
    pages = "031",
    year = "2008"
}

@article{Poland:2018epd,
    author = "Poland, David and Rychkov, Slava and Vichi, Alessandro",
    title = "{The Conformal Bootstrap: Theory, Numerical Techniques, and Applications}",
    eprint = "1805.04405",
    archivePrefix = "arXiv",
    primaryClass = "hep-th",
    doi = "10.1103/RevModPhys.91.015002",
    journal = "Rev. Mod. Phys.",
    volume = "91",
    number = "1",
    pages = "015002",
    year = "2019"
}

@article{El-Showk:2012cjh,
    author = "El-Showk, Sheer and Paulos, Miguel F. and Poland, David and Rychkov, Slava and Simmons-Duffin, David and Vichi, Alessandro",
    title = "{Solving the 3D Ising Model with the Conformal Bootstrap}",
    eprint = "1203.6064",
    archivePrefix = "arXiv",
    primaryClass = "hep-th",
    doi = "10.1103/PhysRevD.86.025022",
    journal = "Phys. Rev. D",
    volume = "86",
    pages = "025022",
    year = "2012"
}

@article{El-Showk:2014dwa,
    author = "El-Showk, Sheer and Paulos, Miguel F. and Poland, David and Rychkov, Slava and Simmons-Duffin, David and Vichi, Alessandro",
    title = "{Solving the 3d Ising Model with the Conformal Bootstrap II. $c$-Minimization and Precise Critical Exponents}",
    eprint = "1403.4545",
    archivePrefix = "arXiv",
    primaryClass = "hep-th",
    doi = "10.1007/s10955-014-1042-7",
    journal = "J. Stat. Phys.",
    volume = "157",
    pages = "869",
    year = "2014"
}

@article{Sachdev:1992fk,
    author = "Sachdev, Subir and Ye, Jinwu",
    title = "{Gapless spin fluid ground state in a random, quantum Heisenberg magnet}",
    eprint = "cond-mat/9212030",
    archivePrefix = "arXiv",
    doi = "10.1103/PhysRevLett.70.3339",
    journal = "Phys. Rev. Lett.",
    volume = "70",
    pages = "3339",
    year = "1993"
}

@misc{Kitaev:2015sim,
    author = "Kitaev, Alexei",
    title = "{A simple model of quantum holography}",
    howpublished = "Talks at KITP, April 7 and May 27, 2015",
    url = "http://online.kitp.ucsb.edu/online/entangled15/kitaev/",
    year = "2015"
}

@article{Polchinski:2016xgd,
    author = "Polchinski, Joseph and Rosenhaus, Vladimir",
    title = "{The Spectrum in the Sachdev-Ye-Kitaev Model}",
    eprint = "1601.06768",
    archivePrefix = "arXiv",
    primaryClass = "hep-th",
    doi = "10.1007/JHEP04(2016)001",
    journal = "JHEP",
    volume = "04",
    pages = "001",
    year = "2016"
}

@article{Gross:2017hcz,
    author = "Gross, David J. and Rosenhaus, Vladimir",
    title = "{All point correlation functions in SYK}",
    eprint = "1710.08113",
    archivePrefix = "arXiv",
    primaryClass = "hep-th",
    doi = "10.1007/JHEP12(2017)148",
    journal = "JHEP",
    volume = "12",
    pages = "148",
    year = "2017"
}

@article{Aikawa:2025sib,
    author = "Aikawa, Yu and Morita, Takeshi",
    title = "{Bootstrapping Shape Invariance: Numerical Bootstrap as a Detector of Solvable Systems}",
    eprint = "2504.08586",
    archivePrefix = "arXiv",
    primaryClass = "hep-th",
    doi = "10.1093/ptep/ptaf131",
    journal = "PTEP",
    volume = "2025",
    pages = "113A01",
    year = "2025"
}

@article{Guo:2023gfi,
    author = "Guo, Yongwei and Li, Wenliang",
    title = "{Solving anharmonic oscillator with null states: Hamiltonian bootstrap and Dyson-Schwinger equations}",
    eprint = "2305.15992",
    archivePrefix = "arXiv",
    primaryClass = "hep-th",
    doi = "10.1103/PhysRevD.108.125002",
    journal = "Phys. Rev. D",
    volume = "108",
    number = "12",
    pages = "125002",
    year = "2023"
}

@article{Lin:2020mme,
    author = "Lin, Henry W.",
    title = "{Bootstraps to strings: solving random matrix models with positivity}",
    eprint = "2002.08387",
    archivePrefix = "arXiv",
    primaryClass = "hep-th",
    doi = "10.1007/JHEP06(2020)090",
    journal = "JHEP",
    volume = "06",
    pages = "090",
    year = "2020"
}

@article{Bhattacharya:2021btd,
    author = "Bhattacharya, Jyotirmoy and Das, Diptarka and Das, Sayan Kumar and Jha, Ankit Kumar and Kundu, Moulindu",
    title = "{Numerical bootstrap in quantum mechanics}",
    eprint = "2108.11416",
    archivePrefix = "arXiv",
    primaryClass = "hep-th",
    doi = "10.1016/j.physletb.2021.136785",
    journal = "Phys. Lett. B",
    volume = "823",
    pages = "136785",
    year = "2021"
}

@article{Lawrence:2025wyl,
    author = "Lawrence, Scott and McPeak, Brian",
    title = "{Quantum bootstrap for central potentials}",
    eprint = "2512.09041",
    archivePrefix = "arXiv",
    primaryClass = "quant-ph",
    reportNumber = "LA-UR-25-31761",
    month = "12",
    year = "2025"
}

@article{Khan:2022uyz,
    author = "Khan, Sakil and Agarwal, Yuv and Tripathy, Devjyoti and Jain, Sachin",
    title = "{Bootstrapping PT symmetric quantum mechanics}",
    eprint = "2202.05351",
    archivePrefix = "arXiv",
    primaryClass = "quant-ph",
    doi = "10.1016/j.physletb.2022.137445",
    journal = "Phys. Lett. B",
    volume = "834",
    pages = "137445",
    year = "2022"
}

@article{Khan:2024mhc,
    author = "Khan, Sakil and Rathod, Harsh",
    title = "{Bootstrapping non-Hermitian quantum systems}",
    eprint = "2409.06784",
    archivePrefix = "arXiv",
    primaryClass = "hep-th",
    doi = "10.1103/PhysRevD.111.105005",
    journal = "Phys. Rev. D",
    volume = "111",
    number = "10",
    pages = "105005",
    year = "2025"
}

@article{Lin:2023owt,
    author = "Lin, Henry W.",
    title = "{Bootstrap bounds on D0-brane quantum mechanics}",
    eprint = "2302.04416",
    archivePrefix = "arXiv",
    primaryClass = "hep-th",
    doi = "10.1007/JHEP06(2023)038",
    journal = "JHEP",
    volume = "06",
    pages = "038",
    year = "2023"
}

@article{Berenstein:2021dyf,
    author = "Berenstein, David and Hulsey, George",
    title = "{Bootstrapping Simple QM Systems}",
    eprint = "2108.08757",
    archivePrefix = "arXiv",
    primaryClass = "hep-th",
    month = "8",
    year = "2021"
}

@article{Berenstein:2022ygg,
    author = "Berenstein, David and Hulsey, George",
    title = "{Anomalous bootstrap on the half-line}",
    eprint = "2206.01765",
    archivePrefix = "arXiv",
    primaryClass = "hep-th",
    doi = "10.1103/PhysRevD.106.045029",
    journal = "Phys. Rev. D",
    volume = "106",
    number = "4",
    pages = "045029",
    year = "2022"
}

@article{Mccoy:1929com,
	title = {On {Commutation} {Rules} in the {Algebra} of {Quantum} {Mechanics}},
	volume = {15},
	number = {3},
	journal = {Proceedings of the National Academy of Sciences of the United States of America},
	author = {McCoy, Neal H.},
	year = {1929},
	pages = {200--202},
}

@article{Huang:2025sua,
    author = "Huang, Zhijian and Li, Wenliang",
    title = "{Bootstrapping periodic quantum systems}",
    eprint = "2507.02386",
    archivePrefix = "arXiv",
    primaryClass = "hep-th",
    month = "7",
    year = "2025"
}

@article{Vegh:2025kgx,
    author = "Vegh, David",
    title = "{A folded string dual for the Sachdev-Ye-Kitaev model}",
    eprint = "2509.05435",
    archivePrefix = "arXiv",
    primaryClass = "hep-th",
    reportNumber = "QMUL-PH-25-25",
    month = "9",
    year = "2025"
}

@article{Vegh:2024uie,
    author = "Vegh, David",
    title = "{Quantizing the folded string in AdS$_2$}",
    eprint = "2409.06663",
    archivePrefix = "arXiv",
    primaryClass = "hep-th",
    reportNumber = "QMUL-PH-24-13",
    month = "9",
    year = "2024"
}

@article{Al-Hashimi:2020qvi,
    author = "Al-Hashimi, M. H. and Wiese, U. -J.",
    title = "{Alternative momentum concept for a quantum mechanical particle in a box}",
    eprint = "2012.09596",
    archivePrefix = "arXiv",
    primaryClass = "quant-ph",
    doi = "10.1103/PhysRevResearch.3.L042008",
    journal = "Phys. Rev. Res.",
    volume = "3",
    number = "4",
    pages = "L042008",
    year = "2021"
}

@article{Han:2020bkb,
    author = "Han, Xizhi and Hartnoll, Sean A. and Kruthoff, Jorrit",
    title = "{Bootstrapping Matrix Quantum Mechanics}",
    eprint = "2004.10212",
    archivePrefix = "arXiv",
    primaryClass = "hep-th",
    doi = "10.1103/PhysRevLett.125.041601",
    journal = "Phys. Rev. Lett.",
    volume = "125",
    number = "4",
    pages = "041601",
    year = "2020"
}

@article{Maldacena:2016hyu,
    author = "Maldacena, Juan and Stanford, Douglas",
    title = "{Remarks on the Sachdev-Ye-Kitaev model}",
    eprint = "1604.07818",
    archivePrefix = "arXiv",
    primaryClass = "hep-th",
    doi = "10.1103/PhysRevD.94.106002",
    journal = "Phys. Rev. D",
    volume = "94",
    number = "10",
    pages = "106002",
    year = "2016"
}

@article{Sword:2024gvv,
    author = "Sword, Lewis and Vegh, David",
    title = "{Quantum mechanical bootstrap on the interval: Obtaining the exact spectrum}",
    eprint = "2402.03434",
    archivePrefix = "arXiv",
    primaryClass = "hep-th",
    reportNumber = "QMUL-PH-24-02",
    doi = "10.1103/PhysRevD.109.126002",
    journal = "Phys. Rev. D",
    volume = "109",
    number = "12",
    pages = "126002",
    year = "2024"
}

@article{Berenstein:2023ppj,
    author = "Berenstein, David and Hulsey, George",
    title = "{One-dimensional reflection in the quantum mechanical bootstrap}",
    eprint = "2307.11724",
    archivePrefix = "arXiv",
    primaryClass = "hep-th",
    doi = "10.1103/PhysRevD.109.025013",
    journal = "Phys. Rev. D",
    volume = "109",
    number = "2",
    pages = "025013",
    year = "2024"
}

@article{Berenstein:2021loy,
    author = "Berenstein, David and Hulsey, George",
    title = "{Bootstrapping more QM systems}",
    eprint = "2109.06251",
    archivePrefix = "arXiv",
    primaryClass = "hep-th",
    doi = "10.1088/1751-8121/ac7118",
    journal = "J. Phys. A",
    volume = "55",
    number = "27",
    pages = "275304",
    year = "2022"
}

@article{Aikawa:2021eai,
    author = "Aikawa, Yu and Morita, Takeshi and Yoshimura, Kota",
    title = "{Application of bootstrap to a {\ensuremath{\theta}} term}",
    eprint = "2109.02701",
    archivePrefix = "arXiv",
    primaryClass = "hep-th",
    doi = "10.1103/PhysRevD.105.085017",
    journal = "Phys. Rev. D",
    volume = "105",
    number = "8",
    pages = "085017",
    year = "2022"
}
\end{document}